\documentclass[11pt,a4paper]{article}
\usepackage[T1]{fontenc}
\usepackage{lmodern}
\usepackage{microtype}
\usepackage[a4paper,top=2.8cm,bottom=2.8cm,left=3cm,right=3cm]{geometry}
\usepackage{amsmath}
\usepackage{amssymb}
\numberwithin{equation}{section}
\allowdisplaybreaks
\usepackage[backend=biber,style=numeric-comp,sorting=none,doi=true,eprint=true]{biblatex}
\usepackage[hidelinks]{hyperref}
\title{Removing Ostrogradsky modes in multi-field higher-order scalar-tensor theories}
\author{Hamed Bouzari Nezhad\\
\textit{Brussels, Belgium}\\
\href{mailto:hamed.bouzarinezhad@gmail.com}{hamed.bouzarinezhad@gmail.com}}
\date{\today}
\begin{document}
\maketitle
\begin{abstract}
We study how the Ostrogradsky modes can be removed in multi-field higher-order scalar-tensor theories. For a general class of theories with an arbitrary number $\mathcal{N}$ of scalar fields and quadratic dependence on their second derivatives, we perform an ADM decomposition and carry out the Hamiltonian analysis in the branch where the metric kinetic block is invertible. The primary degeneracy condition is a matrix condition in field space. Previous studies of coupled higher-derivative systems have shown that primary degeneracy need not by itself be sufficient. In the generally covariant multi-scalar-tensor setting considered here, where the scalar and metric velocities mix in the ADM decomposition, preservation of the primary degeneracy constraints generates additional consistency conditions. We obtain these conditions explicitly in terms of the ADM coefficient blocks generated by the covariant action. Some of these conditions are antisymmetric in the field-space indices and vanish in the single-field limit. A rank condition on the constraint algebra is required in addition. Under the assumptions used in the Hamiltonian counting, in particular that the scalar second-class sector leaves the diffeomorphism constraints first class, the theory then propagates $2+\mathcal{N}$ degrees of freedom, the two tensor modes of gravity and one scalar mode per field, with no additional Ostrogradsky mode. We check the conditions in the single-field limit and in a multi-field quadratic Horndeski-type subclass. We also give a subclass with nonzero scalar-metric kinetic mixing, constructed directly at the level of the ADM coefficient blocks, showing that the Hamiltonian conditions admit nontrivial solutions at the ADM-block level.
\end{abstract}
\section{Introduction}
\label{Sec: Introduction}
Scalar-tensor theories provide a broad class of extensions of general relativity. They appear in effective descriptions of gravity and in models of inflation and late-time cosmic acceleration. Derivative interactions must be introduced without propagating unwanted degrees of freedom. In the single-field case, the most general scalar-tensor theory with second-order field equations was constructed by Horndeski \cite{Horndeski:1974wa}. Galileons and generalized Galileons organize derivative self-interactions of scalar fields so as to avoid higher-order equations of motion \cite{Nicolis:2008in,Deffayet:2009mn,Deffayet:2011gz,Kobayashi:2011nu}.

Second-order field equations are a sufficient condition for avoiding the Ostrogradsky problem \cite{Woodard:2015zca}, but they are not necessary. Higher-order theories may still propagate the correct number of degrees of freedom when their kinetic structure is degenerate. Kinetic degeneracy gives rise to primary constraints, and an appropriate constraint structure can remove the would-be extra mode from the physical phase space. This observation led to theories beyond Horndeski and to degenerate higher-order scalar-tensor theories \cite{Gleyzes:2014dya,Langlois:2015cwa,Langlois:2015skt,Crisostomi:2016czh,BenAchour:2016cay}. For a single scalar field, the degeneracy conditions for theories quadratic in the second derivatives of the scalar were obtained in \cite{Langlois:2015cwa}. The resulting quadratic degenerate families were later classified and related to Horndeski theory by disformal transformations in \cite{BenAchour:2016cay}. The relevant criterion is therefore the full constraint structure generated by the kinetic degeneracy, rather than the differential order of the field equations alone.

The problem becomes more involved when more than one scalar field is present. In a multi-field theory, the scalar kinetic sector is described by matrices in field space rather than by a single coefficient. Degeneracy is then no longer characterized by a single scalar condition. The null directions of the kinetic matrix may involve different combinations of scalar and metric velocities, and different degeneracy patterns can lead to different constraint structures. Related issues in systems with several higher-derivative fields have been studied previously. The Ostrogradsky problem in the presence of several fields, including gravitational theories, has been discussed in \cite{deRham:2016wji}. For coupled systems with several higher-derivative variables, it was established in \cite{Motohashi:2016ftl,Klein:2016aiq} that degeneracy of the kinetic matrix is in general not sufficient, and that preservation of the resulting primary constraints generates additional conditions. In \cite{Crisostomi:2017aim} the corresponding field-theory analysis was carried out, and the secondary conditions were organized into symmetric and antisymmetric parts in the field-space indices, under assumptions that exclude gauge symmetries and a dynamical spin-2 field.

Multi-field scalar-tensor theories have been studied from several viewpoints. Multi-Galileon theories give derivative interactions involving several scalar fields \cite{Padilla:2010ir,Padilla:2012dx}, and covariant multi-field extensions coupled to gravity were developed in \cite{Kobayashi:2013ina}. The most general multiple-scalar theory in flat space with second-order field equations was obtained in \cite{Sivanesan:2013tba}. Bi-scalar extensions of Horndeski theory have also been classified by imposing second-order field equations \cite{Ohashi:2015fma}, and a systematic Lagrangian realization of the general second-order bi-scalar-tensor field equations has recently been given in \cite{Horndeski:2024hee}. Within the multi-Galileon framework there exist genuinely multi-field operators that have no counterpart in the single-field limit \cite{Allys:2016hfl,Akama:2017jsa}. Multiple scalar-tensor theories have been constructed from interacting hypersurfaces \cite{Yu:2024sed}, where an explicit bi-scalar example is shown to propagate two tensor and two scalar degrees of freedom at the level of linear cosmological perturbations, rather than through a general nonlinear Dirac analysis. Explicit two-field Horndeski and DHOST subsets are generated by disformal transformations in \cite{Domenech:2025gao}. These works establish several classes of multi-field scalar-tensor theories. Here we study a different problem: the degeneracy and constraint-consistency conditions of a general quadratic higher-order multi-field theory. We do not take second-order field equations as the defining principle, and we do not proceed by constructing particular models. Instead, we consider a multi-field scalar-tensor theory containing terms quadratic in the second covariant derivatives of the scalar fields, and study the degeneracy and consistency conditions required to remove the unwanted higher-order modes.

The aim of this paper is to formulate and analyze a general quadratic multi-field higher-order scalar-tensor theory in a form where its Hamiltonian constraint structure can be followed explicitly. We introduce the covariant action and its independent building blocks, and then perform an ADM decomposition \cite{Gourgoulhon:2007ue}. To isolate the kinetic structure, we use an auxiliary-field formulation, which makes the velocities of the scalar and metric variables manifest. The corresponding Hessian has a block structure, with scalar, metric, and mixed scalar-metric components, and we use this structure to classify the possible primary degeneracy branches.

Among these branches, we focus on the case in which the metric kinetic block is invertible. In this branch the metric kinetic sector is nondegenerate, and the degeneracy arises through the mixing between scalar and metric velocities. The primary degeneracy condition is then a matrix condition in field space, and it leads to primary constraints associated with the higher-order scalar sector. We follow the time evolution of these constraints in the Hamiltonian formalism \cite{Dirac:1964,Henneaux:1992ig}. For more than one field, preservation of the primary constraints generates further consistency conditions, and some of these are antisymmetric in the field-space indices and so have no analogue in the single-field theory. As recalled above, this mechanism is known in general higher-derivative systems. The present work realizes this mechanism in a generally covariant multi-scalar-tensor theory with a dynamical metric. The scalar and metric velocities mix in the ADM decomposition, the gravitational constraints are present, and the number of fields $\mathcal{N}$ is arbitrary. The consistency conditions are obtained explicitly in terms of the ADM coefficient blocks generated by the covariant action.

The main result is a set of conditions in the metric-invertible branch which, together with the assumptions stated in subsection \ref{Subsec: Counting degrees of freedom}, give a constraint structure in which the quadratic multi-field theory propagates no Ostrogradsky modes associated with the higher-order scalar sector. These conditions are the primary degeneracy condition, the primary-primary consistency conditions generated by preserving the primary constraints, and a final rank condition ensuring that the constraint algorithm closes in the desired way. The assumptions of subsection \ref{Subsec: Counting degrees of freedom} include, in particular, that the scalar second-class sector leaves the diffeomorphism constraints first class. Under these conditions and assumptions, the theory propagates $2+\mathcal{N}$ degrees of freedom, the two tensor modes of gravity together with one scalar mode for each of the $\mathcal{N}$ scalar fields. The analysis does not provide a complete classification of healthy multi-field theories.

The paper is organized as follows. In section \ref{Sec: Covariant action and auxiliary field formulation} we introduce the multi-field quadratic action, define the relevant covariant building blocks and rewrite the theory in auxiliary-field form. In section \ref{Sec: ADM decomposition of the action} we perform the ADM decomposition and identify the coefficient blocks controlling the velocity sector. In section \ref{Sec: Primary degeneracy conditions} we formulate the primary degeneracy conditions and distinguish the possible degeneracy branches. In section \ref{Sec: Complete degeneracy in the metric-invertible branch} we focus on the metric-invertible branch and derive the complete primary degeneracy condition. In section \ref{Sec: Hamiltonian analysis} we carry out the Hamiltonian analysis, derive the primary-primary consistency conditions, obtain the secondary constraints and give the degree-of-freedom count. In section \ref{Sec: Consistency checks and special subclasses} we discuss the single-field limit, a multi-field Horndeski-type subclass and a constructive multi-field subclass satisfying these conditions; the last of these is constructed directly at the level of the ADM coefficient blocks, and we do not solve the inverse map back to explicit covariant coefficient functions $A^{(I)}$, so it should be understood as an ADM-level subclass. We discuss our results and conclude in section \ref{Sec: Discussion and conclusion}. The appendices collect several lengthy coefficient expressions and give further details of the constraint analysis.
\section{Covariant action and auxiliary field formulation}
\label{Sec: Covariant action and auxiliary field formulation}
We consider a theory of $\mathcal{N}$ scalar fields $\phi^{i}$, with $i=1,\cdots,\mathcal{N}$, which generalizes the single-field quadratic higher-order scalar-tensor construction by promoting the scalar $\phi$ to a collection $\phi^{i}$. Field-space indices are denoted by Latin letters $i,j,k,\cdots$, while spacetime indices are denoted by Greek letters $\mu,\nu,\rho,\cdots$ and are raised and lowered with $g_{\mu\nu}$. The first and second covariant derivatives of the scalar fields are
\begin{align}
\label{Eq: Scalar derivatives}
\phi^{i}{}_{\mu}\equiv\nabla_{\mu}\phi^{i},\qquad
\phi^{i}{}_{\mu\nu}\equiv\nabla_{\mu}\nabla_{\nu}\phi^{i},
\end{align}
where $\nabla_{\mu}$ is the Levi-Civita covariant derivative of $g_{\mu\nu}$. Since $\nabla_{\mu}$ is torsion-free, the scalar Hessian $\phi^{i}{}_{\mu\nu}$ is symmetric, $\phi^{i}{}_{\mu\nu}=\phi^{i}{}_{(\mu\nu)}$. The kinetic matrix in field space is
\begin{align}
\label{Eq: Kinetic matrix}
X^{ij}=g^{\mu\nu}\phi^{i}{}_{\mu}\phi^{j}{}_{\nu},
\end{align}
which is symmetric, $X^{ij}=X^{(ij)}$. The coefficient functions of the theory depend on the scalar fields $\phi^{i}$ and on the independent components of $X^{ij}$; an argument such as $(\phi^{i},X^{ij})$ is understood schematically as dependence on all $\mathcal{N}$ fields and on all independent components of the symmetric kinetic matrix.

The interactions are built from the metric, the first derivatives $\phi^{i}{}_{\mu}$, and the scalar Hessian $\phi^{i}{}_{\mu\nu}$, organized by their power of $\phi^{i}{}_{\mu\nu}$. At linear order in the Hessian there are two parity-even structures: the trace
\begin{align}
\label{Eq: Trace of scalar Hessian}
\phi^{i}{}_{\mu}{}^{\mu}=g^{\mu\nu}\phi^{i}{}_{\mu\nu},
\end{align}
and the contraction with two first derivatives, $\phi^{i\mu}\phi^{j\nu}\phi^{k}{}_{\mu\nu}$. At quadratic order in the Hessian the parity-even contractions built from $g_{\mu\nu}$ and $\phi^{i}{}_{\mu}$ are
\begin{align}
\label{Eq: Quadratic blocks}
&L_{1}^{ij}=\phi^{i}{}_{\mu\nu}\phi^{j\mu\nu},\qquad
L_{2}^{ij}=\phi^{i}{}_{\mu}{}^{\mu}\phi^{j}{}_{\nu}{}^{\nu},\qquad
L_{3}^{ijkl}=\phi^{i\mu}\phi^{j\nu}\phi^{k}{}_{\mu\nu}\phi^{l}{}_{\rho}{}^{\rho},\nonumber\\&
L_{4}^{ijkl}=\phi^{i\mu}\phi^{j\nu}\phi^{k}{}_{\mu\rho}\phi^{l}{}_{\nu}{}^{\rho},\qquad
L_{5}^{ijklmn}=\phi^{i\mu}\phi^{j\nu}\phi^{k\rho}\phi^{l\sigma}\phi^{m}{}_{\mu\nu}\phi^{n}{}_{\rho\sigma}.
\end{align}

These five blocks are obtained by classifying the scalars quadratic in $\phi^{i}{}_{\mu\nu}$ according to the number of first-derivative factors they carry: none for $L_{1}^{ij}$ and $L_{2}^{ij}$, two for $L_{3}^{ijkl}$ and $L_{4}^{ijkl}$, and four for $L_{5}^{ijklmn}$. They form the complete parity-even basis quadratic in $\phi^{i}{}_{\mu\nu}$ within the sector built from $g_{\mu\nu}$ and $\phi^{i}{}_{\mu}$ with no explicit curvature tensor. Further contractions of first derivatives with one another carry no Hessian factor and are absorbed into the arbitrary $X^{ij}$-dependence of the coefficient functions, so they generate no independent structures.

The symmetry of the Hessian induces field-space symmetries of these blocks. The blocks $L_{1}^{ij}$ and $L_{2}^{ij}$ are symmetric under $i\leftrightarrow j$. The block $L_{3}^{ijkl}$ is symmetric under $i\leftrightarrow j$, since $\phi^{k}{}_{\mu\nu}$ is symmetric in its spacetime indices. The block $L_{4}^{ijkl}$ is invariant under the simultaneous exchange $(i,k)\leftrightarrow(j,l)$, and $L_{5}^{ijklmn}$ is invariant under $i\leftrightarrow j$, under $k\leftrightarrow l$, and under the simultaneous exchange $(i,j,m)\leftrightarrow(k,l,n)$ of its two linear-Hessian factors. The linear structure $\phi^{i\mu}\phi^{j\nu}\phi^{k}{}_{\mu\nu}$ is symmetric under $i\leftrightarrow j$.

Collecting these structures together with the non-minimal curvature coupling, the action is
\begin{align}
\label{Eq: Action}
S&=\int d^{4}x\sqrt{-g}\Big[F(\phi^{i},X^{ij})R+P(\phi^{i},X^{ij})+Q_{i}(\phi^{i},X^{ij})\phi^{i}{}_{\mu}{}^{\mu}+B_{ijk}(\phi^{i},X^{ij})\phi^{i\mu}\phi^{j\nu}\phi^{k}{}_{\mu\nu}\nonumber\\&+A^{(1)}_{ij}(\phi^{i},X^{ij})L_{1}^{ij}+A^{(2)}_{ij}(\phi^{i},X^{ij})L_{2}^{ij}+A^{(3)}_{ijkl}(\phi^{i},X^{ij})L_{3}^{ijkl}+A^{(4)}_{ijkl}(\phi^{i},X^{ij})L_{4}^{ijkl}\nonumber\\&+A^{(5)}_{ijklmn}(\phi^{i},X^{ij})L_{5}^{ijklmn}\Big],
\end{align}
where $R$ is the Ricci scalar of $g_{\mu\nu}$ and the coefficient functions $F$, $P$, $Q_{i}$, $B_{ijk}$ and $A^{(I)}$ are arbitrary functions of $\phi^{i}$ and $X^{ij}$, subject only to the field-space symmetries below. Because each block contributes to the action only through the part of its coefficient that shares its symmetry, we may impose without loss of generality
\begin{align}
\label{Eq: Symmetries of the coefficients}
&B_{ijk}=B_{jik},\qquad
A^{(1)}_{ij}=A^{(1)}_{ji},\qquad
A^{(2)}_{ij}=A^{(2)}_{ji},\qquad
A^{(3)}_{ijkl}=A^{(3)}_{jikl},\qquad
A^{(4)}_{ijkl}=A^{(4)}_{jilk},\nonumber\\&
A^{(5)}_{ijklmn}=A^{(5)}_{jiklmn}=A^{(5)}_{ijlkmn}=A^{(5)}_{klijnm}.
\end{align}

The antisymmetric parts of the coefficients with respect to these transformations multiply identically symmetric blocks and drop out of the action.\footnote{If $\mathsf{n}(\cdot)$ denotes the number of independent field-space components, the symmetries in Eq. \eqref{Eq: Symmetries of the coefficients} give
\begin{align}
\label{Eq: Number of independent components}
&\mathsf{n}\left({B_{ijk}}\right)=\frac{\mathcal{N}^{2}(\mathcal{N}+1)}{2},\quad
\mathsf{n}\left({A^{(1)}_{ij}}\right)=\frac{\mathcal{N}(\mathcal{N}+1)}{2},\quad
\mathsf{n}\left({A^{(2)}_{ij}}\right)=\frac{\mathcal{N}(\mathcal{N}+1)}{2},\quad
\mathsf{n}\left({A^{(3)}_{ijkl}}\right)=\frac{\mathcal{N}^{3}(\mathcal{N}+1)}{2},\nonumber\\&
\mathsf{n}\left({A^{(4)}_{ijkl}}\right)=\frac{\mathcal{N}^{4}+\mathcal{N}^{2}}{2},\quad
\mathsf{n}\left({A^{(5)}_{ijklmn}}\right)=\frac{\mathcal{N}^{6}+2\mathcal{N}^{5}+\mathcal{N}^{4}+2\mathcal{N}^{3}+2\mathcal{N}^{2}}{8}.
\end{align}}

Equation \eqref{Eq: Action} is written in a reduced parity-even basis, in which the only explicit curvature term is the coupling $FR$. Curvature-gradient operators constructed from the Ricci tensor, the ordinary Riemann tensor, or the double-dual Riemann tensor are not added as independent terms. For example, using the Ricci identity and integrating by parts, one obtains
\begin{align}
\label{Eq: Reduction of Ricci coupling}
C_{ij}R_{\mu\nu}\phi^{i\mu}\phi^{j\nu}&\simeq C_{ij,\phi^{k}}X^{jk}\phi^{i}{}_{\mu}{}^{\mu}-C_{ik,\phi^{j}}\phi^{i\mu}\phi^{j\nu}\phi^{k}{}_{\mu\nu}+C_{ij}\left(L_{2}^{ij}-L_{1}^{ij}\right)\nonumber\\&+2C_{ik,X^{jl}}\left(L_{3}^{ijlk}-L_{4}^{ijkl}\right),
\end{align}
where, here and in the following, $\simeq$ denotes equality up to boundary terms and
\begin{align}
\label{Eq: Derivatives of Ricci coefficient}
C_{ij,\phi^{k}}=\frac{\partial C_{ij}}{\partial\phi^{k}},\qquad
C_{ij,X^{kl}}=\frac{\partial C_{ij}}{\partial X^{kl}}.
\end{align}

Since $R_{\mu\nu}$ is symmetric, only the symmetric part of $C_{ij}$ contributes to the left-hand side of Eq. \eqref{Eq: Reduction of Ricci coupling}. Moreover, since $X^{ij}=X^{ji}$, we take $C_{ij,X^{kl}}=C_{ij,X^{lk}}$. The four terms on the right-hand side of Eq. \eqref{Eq: Reduction of Ricci coupling} are already contained in the present basis. The first term has the trace-linear form $Q_{i}\phi^{i}{}_{\mu}{}^{\mu}$. The second term has the linear Hessian form $B_{ijk}\phi^{i\mu}\phi^{j\nu}\phi^{k}{}_{\mu\nu}$, after symmetrizing its coefficient in the two first-derivative labels. The third term belongs to the quadratic blocks $L_{1}^{ij}$ and $L_{2}^{ij}$, while the last term belongs to the quadratic blocks $L_{3}^{ijkl}$ and $L_{4}^{ijkl}$. Thus the Ricci coupling does not have to be included independently.

Similarly, ordinary Riemann-gradient couplings of the form
\begin{align}
\label{Eq: Riemann gradient coupling}
D_{ijkl}R_{\alpha\beta\gamma\delta}\phi^{i\alpha}\phi^{j\beta}\phi^{k\gamma}\phi^{l\delta},
\end{align}
are reducible, up to integration by parts and coefficient redefinitions, to structures already present in Eq. \eqref{Eq: Action}. In particular, their reduction gives terms belonging to the linear Hessian sector $B_{ijk}\phi^{i\mu}\phi^{j\nu}\phi^{k}{}_{\mu\nu}$ and to the quadratic blocks $L_{3}^{ijkl}$, $L_{4}^{ijkl}$ and $L_{5}^{ijklmn}$.

Parity-even contractions involving the double-dual Riemann tensor do not define an independent sector in four dimensions. With our convention
\begin{align}
\label{Eq: Double dual Riemann}
{}^{*}R^{*}_{\mu\nu\rho\sigma}\equiv\frac{1}{4}\epsilon_{\mu\nu}{}^{\alpha\beta}R_{\alpha\beta\gamma\delta}\epsilon^{\gamma\delta}{}_{\rho\sigma},
\end{align}
one has
\begin{align}
\label{Eq: Double dual Riemann decomposition}
{}^{*}\!R^{*}_{\mu\nu\rho\sigma}=-R_{\mu\nu\rho\sigma}-Rg_{[\mu|\rho|}g_{\nu]\sigma}+2g_{[\mu|\rho|}R_{\nu]\sigma}-2g_{[\mu|\sigma|}R_{\nu]\rho}.
\end{align}

Therefore its ordinary Riemann part reduces to the $B_{ijk}$ sector and to the blocks $L_{3}^{ijkl}$, $L_{4}^{ijkl}$ and $L_{5}^{ijklmn}$, while its Ricci part reduces according to Eq. \eqref{Eq: Reduction of Ricci coupling}. Its Ricci scalar part is absorbed into the non-minimal coupling $FR$, since the remaining contractions of first derivatives give functions of $\phi^{i}$ and $X^{ij}$. Parity-odd couplings involving a single dual Riemann tensor are not considered in the present parity-even action.

For the Hamiltonian analysis it is convenient to remove the higher derivatives of the scalar fields by introducing an auxiliary co-vector field $\mathcal{A}^{i}{}_{\mu}$ for each scalar, with the relation between $\mathcal{A}^{i}{}_{\mu}$ and $\phi^{i}$ enforced by a Lagrange multiplier. In the auxiliary formulation we replace
\begin{align}
\label{Eq: Auxiliary replacements}
\phi^{i}{}_{\mu}\rightarrow\mathcal{A}^{i}{}_{\mu},\qquad
\phi^{i}{}_{\mu\nu}\rightarrow\mathcal{A}^{i}{}_{\mu\nu}\equiv\nabla_{(\mu}\mathcal{A}^{i}{}_{\nu)},\qquad
X^{ij}\rightarrow g^{\mu\nu}\mathcal{A}^{i}{}_{\mu}\mathcal{A}^{j}{}_{\nu}.
\end{align}

The symmetrized derivative $\mathcal{A}^{i}{}_{\mu\nu}$ reflects the symmetry of the original scalar Hessian $\phi^{i}{}_{\mu\nu}$. Off shell $\mathcal{A}^{i}{}_{\mu}$ is an independent field, so $\nabla_{\mu}\mathcal{A}^{i}{}_{\nu}$ is not symmetric in general; on shell $\mathcal{A}^{i}{}_{\mu}$ is constrained to the gradient of $\phi^{i}$. The Lagrange multiplier contribution is
\begin{align}
\label{Eq: Lagrange multiplier}
S_{\lambda}=\int d^{4}x\sqrt{-g}\lambda_{i}{}^{\mu}\left(\nabla_{\mu}\phi^{i}-\mathcal{A}^{i}{}_{\mu}\right).
\end{align}

Variation with respect to $\lambda_{i}{}^{\mu}$ imposes
\begin{align}
\label{Eq: Auxiliary constraint}
\mathcal{A}^{i}{}_{\mu}=\nabla_{\mu}\phi^{i}.
\end{align}

On the constraint surface one then has
\begin{align}
\label{Eq: Auxiliary Hessian on shell}
\mathcal{A}^{i}{}_{\mu\nu}=\nabla_{(\mu}\mathcal{A}^{i}{}_{\nu)}=\nabla_{\mu}\nabla_{\nu}\phi^{i}=\phi^{i}{}_{\mu\nu},
\end{align}
and the auxiliary expression for $X^{ij}$ reduces to Eq. \eqref{Eq: Kinetic matrix}. Hence the first-order action obtained from Eq. \eqref{Eq: Action} by the replacements in Eq. \eqref{Eq: Auxiliary replacements}, together with $S_{\lambda}$, is equivalent to the original higher-order action, and provides the starting point for the ADM decomposition and Hamiltonian analysis of the following sections.
\section{ADM decomposition of the action}
\label{Sec: ADM decomposition of the action}
We now rewrite the covariant action in a form suitable for the Hamiltonian analysis, using the standard ADM decomposition \cite{Gourgoulhon:2007ue}. We foliate spacetime by spacelike hypersurfaces $\Sigma_{t}$ labelled by a time function $t$ and choose coordinates $x^{\mu}=(t,y^{a})$ adapted to this foliation, where $y^{a}$ are coordinates on $\Sigma_{t}$. Spatial indices on $\Sigma_{t}$ are denoted by Latin letters $a,b,c,d,e,f$. In these adapted coordinates, the line element is written as
\begin{align}
\label{Eq: ADM metric}
ds^{2}=-N^{2}dt^{2}+h_{ab}(dy^{a}+N^{a}dt)(dy^{b}+N^{b}dt),
\end{align}
where $N$ is the lapse, $N^{a}$ is the shift vector, and $h_{ab}$ is the induced metric on $\Sigma_{t}$. The future-directed unit normal satisfies $n_{\mu}n^{\mu}=-1$ and, in the adapted coordinates above, is given by
\begin{align}
\label{Eq: Normal vector in ADM variables}
n_{\mu}=(-N,0,0,0),\qquad
n^{\mu}=\frac{1}{N}(1,-N^{a}).
\end{align}

The time-flow vector is decomposed as
\begin{align}
\label{Eq: Time-flow vector}
t^{\mu}=Nn^{\mu}+e^{\mu}{}_{a}N^{a},
\end{align}
where $e^{\mu}{}_{a}$ are tangent basis vectors on $\Sigma_{t}$. They convert spatial tensors into spacetime tensors, while their duals project spacetime tensors onto the hypersurfaces. The induced metric is
\begin{align}
\label{Eq: Induced metric}
h_{ab}=e^{\mu}{}_{a}e^{\nu}{}_{b}g_{\mu\nu},
\end{align}
and its spacetime form is $h_{\mu\nu}=g_{\mu\nu}+n_{\mu}n_{\nu}$, or equivalently
\begin{align}
\label{Eq: Metric decomposition}
g_{\mu\nu}=h_{\mu\nu}-n_{\mu}n_{\nu},\qquad
h_{\mu}{}^{\nu}=\delta_{\mu}{}^{\nu}+n_{\mu}n^{\nu}.
\end{align}

Tensors fully projected by $e^{\mu}{}_{a}$ are intrinsic to $\Sigma_{t}$. Their spatial covariant derivative is denoted by $D_{a}$ and is compatible with the induced metric, $D_{a}h_{bc}=0$.

The derivative of the normal vector is decomposed as
\begin{align}
\label{Eq: Extrinsic curvature convention}
\nabla_{\mu}n_{\nu}=K_{\mu\nu}-n_{\mu}a_{\nu},
\end{align}
where $K_{\mu\nu}$ is the extrinsic curvature of $\Sigma_{t}$ and is tangent to the hypersurfaces in both indices. Its spatial components are $K_{ab}=e^{\mu}{}_{a}e^{\nu}{}_{b}K_{\mu\nu}$, and $K=h^{ab}K_{ab}$. Contracting Eq. \eqref{Eq: Extrinsic curvature convention} with $n^{\mu}$ gives the acceleration of the normal congruence,
\begin{align}
\label{Eq: Acceleration}
a_{\mu}=n^{\nu}\nabla_{\nu}n_{\mu},\qquad
a_{a}=e^{\mu}{}_{a}a_{\mu}=D_{a}\ln{N}.
\end{align}

The Ricci scalar is decomposed by the Gauss-Codazzi relation. In the above convention,
\begin{align}
\label{Eq: Gauss-Codazzi}
R={}^{(3)}\!R+K_{ab}K^{ab}-K^{2}+2\nabla_{\mu}\left(Kn^{\mu}-a^{\mu}\right),
\end{align}
where ${}^{(3)}\!R$ is the Ricci scalar of $h_{ab}$. The last term in Eq. \eqref{Eq: Gauss-Codazzi} contributes nontrivially because $F$ depends on $\phi^{i}$ and $X^{ij}$. After introducing the auxiliary field and using the auxiliary form of $X^{ij}$ from Eq. \eqref{Eq: Auxiliary replacements}, integration by parts gives
\begin{align}
2\int d^{4}x\sqrt{-g}F\nabla_{\mu}(Kn^{\mu}-a^{\mu})&\simeq2\int d^{4}x\sqrt{-g}(F_{\phi^{i}}\mathcal{A}^{i}{}_{\mu}+2F_{X^{ij}}\mathcal{A}^{i\nu}\nabla_{\mu}\mathcal{A}^{j}{}_{\nu})(a^{\mu}-Kn^{\mu}),\nonumber\\&
\label{Eq: Integration by parts of Gauss-Codazzi boundary term}
\end{align}
where
\begin{align}
\label{Eq: Derivatives of nonminimal coefficient}
F_{\phi^{i}}=\frac{\partial F}{\partial\phi^{i}},\qquad
F_{X^{ij}}=\frac{\partial F}{\partial X^{ij}}.
\end{align}

Since $X^{ij}$ is symmetric by Eq. \eqref{Eq: Kinetic matrix}, $F_{X^{ij}}$ is symmetric in $i,j$. We next decompose the scalar and auxiliary variables into their normal and tangential parts. For the scalar fields,
\begin{align}
\label{Eq: Decomposition of scalar derivative}
\nabla_{\mu}\phi^{i}=e_{\mu}{}^{a}D_{a}\phi^{i}-n_{\mu}\phi_{*}^{i},\qquad
\phi_{*}^{i}=n^{\mu}\nabla_{\mu}\phi^{i}.
\end{align}

For the auxiliary co-vector,
\begin{align}
\label{Eq: Decomposition of auxiliary covector}
\mathcal{A}^{i}{}_{\mu}=e_{\mu}{}^{a}\hat{\mathcal{A}}^{i}{}_{a}-n_{\mu}\mathcal{A}_{*}^{i},\qquad
\hat{\mathcal{A}}^{i}{}_{a}=e^{\mu}{}_{a}\mathcal{A}^{i}{}_{\mu},\qquad
\mathcal{A}_{*}^{i}=n^{\mu}\mathcal{A}^{i}{}_{\mu}.
\end{align}

The spatial part of the auxiliary constraint, enforced by the corresponding component of the Lagrange multiplier $\lambda_{i}{}^{\mu}$, identifies $\hat{\mathcal{A}}^{i}{}_{a}$ with $D_{a}\phi^{i}$. Since this constraint contains no velocities, it can be used before extracting the kinetic structure of the ADM action. With this understanding, no independent normal derivative of $\hat{\mathcal{A}}^{i}{}_{a}$ appears, and the symmetrized derivative $\mathcal{A}^{i}{}_{\mu\nu}$ decomposes as
\begin{align}
\label{Eq: Decomposition of auxiliary Hessian}
\mathcal{A}^{i}{}_{\mu\nu}=n_{\mu}n_{\nu}\mathcal{V}_{*}^{i}-2n_{(\mu}e_{\nu)}{}^{a}(D_{a}\mathcal{A}_{*}^{i}-\hat{\mathcal{A}}^{ib}K_{ab})+e_{\mu}{}^{a}e_{\nu}{}^{b}(S^{i}{}_{ab}-\mathcal{A}_{*}^{i}K_{ab}),
\end{align}
where
\begin{align}
\label{Eq: Definition of auxiliary ADM variables}
\mathcal{V}_{*}^{i}=n^{\mu}n^{\nu}\mathcal{A}^{i}{}_{\mu\nu},\qquad
S^{i}{}_{ab}=D_{(a}\hat{\mathcal{A}}^{i}{}_{b)}.
\end{align}

Only the symmetric spatial derivative $S^{i}{}_{ab}$ appears because the covariant action depends on $\mathcal{A}^{i}{}_{\mu\nu}$, which is symmetric in its spacetime indices. Contracting Eq. \eqref{Eq: Decomposition of auxiliary Hessian} with $g^{\mu\nu}$ gives
\begin{align}
\label{Eq: Trace of auxiliary Hessian}
g^{\mu\nu}\mathcal{A}^{i}{}_{\mu\nu}=-\mathcal{V}_{*}^{i}+S^{i}-\mathcal{A}_{*}^{i}K,
\end{align}
where $S^{i}=h^{ab}S^{i}{}_{ab}$. Using Eq. \eqref{Eq: Decomposition of auxiliary covector}, the kinetic matrix decomposes as
\begin{align}
\label{Eq: Decomposition of kinetic matrix}
X^{ij}=\hat{\mathcal{A}}^{ij}-\mathcal{A}_{*}^{i}\mathcal{A}_{*}^{j},\qquad
\hat{\mathcal{A}}^{ij}=\hat{\mathcal{A}}^{i}{}_{a}\hat{\mathcal{A}}^{ja}.
\end{align}

Finally, the Lagrange multiplier is decomposed as
\begin{align}
\label{Eq: Decomposition of Lagrange multiplier}
\lambda_{i}{}^{\mu}=e^{\mu}{}_{a}\hat{\lambda}_{i}{}^{a}-n^{\mu}\lambda_{*i},\qquad
\hat{\lambda}_{i}{}^{a}=e_{\mu}{}^{a}\lambda_{i}{}^{\mu},\qquad
\lambda_{*i}=n_{\mu}\lambda_{i}{}^{\mu}.
\end{align}

With this sign convention, the multiplier sector gives $\lambda_{*i}(\mathcal{A}_{*}^{i}-\phi_{*}^{i})+\hat{\lambda}_{i}{}^{a}(D_{a}\phi^{i}-\hat{\mathcal{A}}^{i}{}_{a})$.

Putting these decompositions together, the covariant action takes the ADM form
\begin{align}
\label{Eq: ADM action}
S_{\rm{ADM}}&=\int dtd^{3}y\sqrt{h}N\Big[
F{}^{(3)}\!R
+a_{a}\mathcal{E}^{a}
+\mathcal{U}
+\mathcal{C}_{i}\mathcal{V}_{*}^{i}
+\mathcal{M}^{ab}K_{ab}
+\mathcal{G}_{ij}\mathcal{V}_{*}^{i}\mathcal{V}_{*}^{j}\nonumber\\&
+2\mathcal{B}_{i}{}^{ab}\mathcal{V}_{*}^{i}K_{ab}
+\mathcal{K}^{abcd}K_{ab}K_{cd}
+\lambda_{*i}(\mathcal{A}_{*}^{i}-\phi_{*}^{i})
+\hat{\lambda}_{i}{}^{a}(D_{a}\phi^{i}-\hat{\mathcal{A}}^{i}{}_{a})
\Big].
\end{align}

The quantities $\mathcal{V}_{*}^{i}$ and $K_{ab}$ contain the velocities of $\mathcal{A}_{*}^{i}$ and $h_{ab}$, respectively, and will be the relevant velocity variables in the Hamiltonian analysis. The explicit expressions for the ADM coefficients $\mathcal{E}^{a}$, $\mathcal{U}$, $\mathcal{C}_{i}$, $\mathcal{M}^{ab}$, $\mathcal{G}_{ij}$, $\mathcal{B}_{i}{}^{ab}$ and $\mathcal{K}^{abcd}$ are lengthy and are collected in appendix \ref{AppSec: Explicit coefficients in the ADM action}. Their index symmetries are
\begin{align}
\label{Eq: ADM coefficient symmetries}
\mathcal{G}_{ij}=\mathcal{G}_{ji},\qquad
\mathcal{M}^{ab}=\mathcal{M}^{ba},\qquad
\mathcal{B}_{i}{}^{ab}=\mathcal{B}_{i}{}^{ba},\qquad
\mathcal{K}^{abcd}=\mathcal{K}^{bacd}=\mathcal{K}^{abdc}=\mathcal{K}^{cdab}.
\end{align}

All acceleration dependence in Eq. \eqref{Eq: ADM action} is encoded in $a_{a}\mathcal{E}^{a}$. Since $a_{a}=D_{a}\ln{N}$, this term can be rewritten by a spatial integration by parts as
\begin{align}
\label{Eq: Integration by parts of acceleration term}
\int dtd^{3}y\sqrt{h}Na_{a}\mathcal{E}^{a}\simeq-\int dtd^{3}y\sqrt{h}ND_{a}\mathcal{E}^{a}.
\end{align}

Thus, up to a boundary term, the acceleration contribution can be traded for a term without explicit dependence on $a_{a}$.
\section{Primary degeneracy conditions}
\label{Sec: Primary degeneracy conditions}
\subsection{Velocity Hessian and primary constraints}
\label{Subsec: Velocity Hessian and primary constraints}
We now derive the primary degeneracy conditions. The relevant object is the Hessian of the Lagrangian density with respect to the variables that contain time derivatives. In the ADM action \eqref{Eq: ADM action}, these variables are $\mathcal{V}_{*}^{i}$ and $K_{ab}$. As discussed in section \ref{Sec: ADM decomposition of the action}, the spatial components $\hat{\mathcal{A}}^{i}{}_{a}$ are fixed by their own constraint and carry no independent velocity. Therefore the velocity space relevant for the primary degeneracy analysis is spanned only by $\mathcal{V}_{*}^{i}$ and $K_{ab}$. Terms that are independent of these variables, or only linear in them, do not contribute to this Hessian. They can shift the canonical momenta and affect the explicit form of the constraints, the Hamiltonian and the secondary constraint analysis, but they do not affect the rank of the velocity Hessian. Therefore the primary degeneracy condition is determined by the part of the Lagrangian density quadratic in $\mathcal{V}_{*}^{i}$ and $K_{ab}$.

The quadratic velocity sector is
\begin{align}
\label{Eq: Kinetic Lagrangian density}
\mathcal{L}_{\rm{Kin}}=\sqrt{h}N\left(\mathcal{G}_{ij}\mathcal{V}_{*}^{i}\mathcal{V}_{*}^{j}+2\mathcal{B}_{i}{}^{ab}\mathcal{V}_{*}^{i}K_{ab}+\mathcal{K}^{abcd}K_{ab}K_{cd}\right).
\end{align}

We regard this expression as a quadratic form on the space of velocity variables
\begin{align}
\label{Eq: Velocity variables for Hessian}
q^{A}=\left(\mathcal{V}_{*}^{i},K_{ab}\right),
\end{align}
where $K_{ab}=K_{ba}$. The composite index $A$ labels the direct sum of the $\mathcal{N}$-dimensional field-space sector spanned by $\mathcal{V}_{*}^{i}$ and the six-dimensional space of symmetric spatial tensors spanned by $K_{ab}$. Hence the velocity space has dimension $\mathcal{N}+6$.

The primary degeneracy is then determined by the velocity Hessian on this space,
\begingroup
\setlength{\arraycolsep}{0.5em}
\begin{align}
\label{Eq: Hessian}
\mathbb{H}_{AB}=\frac{\partial^{2}\mathcal{L}_{\rm{Kin}}}{\partial q^{A}\partial q^{B}}=2\sqrt{h}N
\begin{pmatrix}
\mathcal{G}_{ij}&\mathcal{B}_{i}{}^{cd}\\
\mathcal{B}_{j}{}^{ab}&\mathcal{K}^{abcd}
\end{pmatrix}.
\end{align}
\endgroup

The overall factor $2\sqrt{h}N$ does not affect the rank of $\mathbb{H}_{AB}$. The Hessian controls the invertibility of the Legendre map from the velocities $\left(\mathcal{V}_{*}^{i},K_{ab}\right)$ to the corresponding canonical momenta. If $\mathbb{H}_{AB}$ is invertible, the velocities can be solved in terms of the momenta and no primary constraint arises from the kinetic sector. If $\mathbb{H}_{AB}$ has null directions, the Legendre map is singular and primary constraints appear.

Let
\begin{align}
\label{Eq: Degeneracy vector}
Z^{A}=\left(\xi^{i},\sigma_{ab}\right),\qquad
\sigma_{ab}=\sigma_{ba},
\end{align}
be a null vector of the Hessian. The condition $\mathbb{H}_{AB}Z^{B}=0$ gives
\begin{align}
\label{Eq: General degeneracy condition I}
&\mathcal{G}_{ij}\xi^{j}+\mathcal{B}_{i}{}^{cd}\sigma_{cd}=0,\nonumber\\&
\mathcal{B}_{j}{}^{ab}\xi^{j}+\mathcal{K}^{abcd}\sigma_{cd}=0.
\end{align}

On a regular region of phase space where the rank of $\mathbb{H}_{AB}$ is constant, the number of primary constraints generated by the velocity degeneracy is
\begin{align}
\label{Eq: General degeneracy condition II}
N_{\rm{Prim}}=\dim\ker\mathbb{H}=\mathcal{N}+6-\operatorname{rank}\mathbb{H}.
\end{align}

This is a primary-level statement. The existence of primary constraints is necessary for removing Ostrogradsky modes, but it is not by itself sufficient. One must also verify that the primary constraints are preserved under time evolution and that the required secondary constraints are generated.
\subsection{Classification of primary degeneracy branches}
\label{Subsec: Classification of primary degeneracy branches}
The system \eqref{Eq: General degeneracy condition I} can be simplified whenever one of the diagonal blocks can be inverted, because one set of null-vector components can then be eliminated in terms of the other. The classification below records which block, if any, is available for this elimination.

The general degeneracy conditions \eqref{Eq: General degeneracy condition I} can be classified according to the invertibility properties of the two diagonal blocks $\mathcal{G}_{ij}$ and $\mathcal{K}^{abcd}$. In what follows, $\det\mathcal{G}$ denotes the determinant of $\mathcal{G}_{ij}$ in field space, while $\det\mathcal{K}$ denotes the determinant of $\mathcal{K}^{abcd}$ as a linear map on the six-dimensional space of symmetric spatial tensors. Similarly, determinants of Schur complements are understood on the corresponding field-space or symmetric-tensor space. With this convention, the following three branches are exhaustive and non-overlapping.
\subsubsection{Branch I: scalar-invertible branch with singular metric sector}
\label{Subsubsec: Branch I: scalar-invertible branch with singular metric sector}
The first branch is characterized by
\begin{align}
\label{Eq: Branch I degeneracy condition I}
\det\mathcal{K}=0,\qquad
\det\mathcal{G}\neq0.
\end{align}

In this case the metric kinetic block is singular, while the scalar block can be inverted. We denote the inverse of $\mathcal{G}_{ij}$ by $\left(\mathcal{G}^{-1}\right)^{ij}$, satisfying
\begin{align}
\label{Eq: Inverse scalar kinetic block}
\mathcal{G}_{ik}\left(\mathcal{G}^{-1}\right)^{kj}=\delta_{i}{}^{j}.
\end{align}

The first equation in \eqref{Eq: General degeneracy condition I} can then be solved for the scalar component of the null vector,
\begin{align}
\label{Eq: Branch I scalar null component}
\xi^{i}=-\left(\mathcal{G}^{-1}\right)^{ij}\mathcal{B}_{j}{}^{ab}\sigma_{ab}.
\end{align}

Substitution into the second equation in \eqref{Eq: General degeneracy condition I} gives
\begin{align}
\label{Eq: Branch I Schur complement equation}
\mathcal{T}^{abcd}\sigma_{cd}=0,
\end{align}
where
\begin{align}
\label{Eq: Branch I Schur complement}
\mathcal{T}^{abcd}=\mathcal{K}^{abcd}-\mathcal{B}_{i}{}^{ab}\left(\mathcal{G}^{-1}\right)^{ij}\mathcal{B}_{j}{}^{cd}.
\end{align}

Thus the primary degeneracy condition in this branch is
\begin{align}
\label{Eq: Branch I degeneracy condition II}
\det\mathcal{T}=0.
\end{align}

The null directions are parameterized by symmetric tensors $\sigma_{ab}\in\ker\mathcal{T}$, and the corresponding scalar components are fixed by Eq. \eqref{Eq: Branch I scalar null component}.

This branch is a legitimate possibility at the level of the primary degeneracy conditions. However, since the metric kinetic block is already singular, the corresponding constraint structure can involve degeneracies intrinsic to the metric sector. Such theories require a separate analysis of the tensor dynamics and of the full Hamiltonian constraint algebra. We therefore do not pursue this branch in the present work.
\subsubsection{Branch II: metric-invertible branch}
\label{Subsubsec: Branch II: metric-invertible branch}
The second branch is characterized by
\begin{align}
\label{Eq: Branch II degeneracy condition I}
\det\mathcal{K}\neq0.
\end{align}

This condition does not impose any invertibility assumption on $\mathcal{G}_{ij}$. It only states that the metric kinetic block can be inverted as a map on symmetric spatial tensors. We denote its inverse by $\left(\mathcal{K}^{-1}\right)_{abcd}$, satisfying
\begin{align}
\label{Eq: Inverse metric kinetic block}
\mathcal{K}^{abef}\left(\mathcal{K}^{-1}\right)_{efcd}=\delta^{a}{}_{(c}\delta^{b}{}_{d)}.
\end{align}

The second equation in \eqref{Eq: General degeneracy condition I} can then be solved for the metric component of the null vector,
\begin{align}
\label{Eq: Branch II metric null component}
\sigma_{ab}=-\left(\mathcal{K}^{-1}\right)_{abcd}\mathcal{B}_{i}{}^{cd}\xi^{i}.
\end{align}

Substituting this into the first equation in \eqref{Eq: General degeneracy condition I} gives
\begin{align}
\label{Eq: Branch II Schur complement equation}
\mathcal{S}_{ij}\xi^{j}=0,
\end{align}
where the effective scalar degeneracy matrix is
\begin{align}
\label{Eq: Branch II Schur complement}
\mathcal{S}_{ij}=\mathcal{G}_{ij}-\mathcal{B}_{i}{}^{ab}\left(\mathcal{K}^{-1}\right)_{abcd}\mathcal{B}_{j}{}^{cd}.
\end{align}

The number of primary constraints in this branch is therefore
\begin{align}
\label{Eq: Number of primary constraints in branch II}
N_{\rm{Prim}}=\dim\ker\mathcal{S}.
\end{align}

Thus $\det\mathcal{S}=0$ guarantees at least one primary constraint, while the number of independent primary constraints is determined by the dimension of the kernel of $\mathcal{S}_{ij}$.

This branch is the one selected for the rest of the analysis. Its defining advantage is that the metric kinetic block is non-degenerate, so the degeneracy is controlled by the effective scalar matrix $\mathcal{S}_{ij}$. The detailed implementation of this condition, including the distinction between partial and complete scalar-sector degeneracy, will be discussed in the next section.
\subsubsection{Branch III: fully singular branch}
\label{Subsubsec: Branch III: fully singular branch}
The third branch is characterized by
\begin{align}
\label{Eq: Branch III degeneracy condition I}
\det\mathcal{K}=0,\qquad
\det\mathcal{G}=0.
\end{align}

In this case neither of the diagonal blocks is invertible, and neither Schur complement is available in general. The degeneracy must therefore be analyzed directly from the coupled system \eqref{Eq: General degeneracy condition I}. Within this branch, different types of null directions can occur.

A pure scalar null direction has the form
\begin{align}
\label{Eq: Branch III pure scalar direction}
\xi^{i}\neq0,\qquad
\sigma_{ab}=0,\qquad
\mathcal{G}_{ij}\xi^{j}=0,\qquad
\mathcal{B}_{i}{}^{ab}\xi^{i}=0.
\end{align}

Such a direction lies in the kernel of the scalar block and is also annihilated by the mixing block. It therefore produces a primary constraint without involving the metric velocity sector.

A pure metric null direction has the form
\begin{align}
\label{Eq: Branch III pure metric direction}
\xi^{i}=0,\qquad
\sigma_{ab}\neq0,\qquad
\mathcal{B}_{i}{}^{ab}\sigma_{ab}=0,\qquad
\mathcal{K}^{abcd}\sigma_{cd}=0.
\end{align}

Such a direction lies in the kernel of the metric block and is also annihilated by the mixing block. It produces a primary constraint intrinsic to the metric velocity sector.

Finally, one may have mixed null directions,
\begin{align}
\label{Eq: Branch III mixed direction}
\xi^{i}\neq0,\qquad
\sigma_{ab}\neq0,
\end{align}
where the pair $\left(\xi^{i},\sigma_{ab}\right)$ solves the full coupled system \eqref{Eq: General degeneracy condition I}. In this case the scalar and metric components of the null vector cannot be disentangled by inverting one of the diagonal blocks. The three cases above are types of null directions within the fully singular branch; a given theory may contain more than one type.

This branch is the most general singular case, but also the least controlled one. Since the metric kinetic block is singular, the theory may contain additional primary constraints in the metric sector, modified tensor dynamics or strongly coupled directions. These possibilities are not excluded by the primary degeneracy analysis alone, but they require a separate Hamiltonian treatment. We therefore leave this branch outside the scope of the present work.
\subsection{Choice of branch and scope of the analysis}
\label{Subsec: Choice of branch and scope of the analysis}
The classification above separates the primary degeneracy problem into three distinct cases. Branch I and Branch III both begin with a singular metric kinetic block and require a separate Hamiltonian analysis. In the remainder of this work we focus on Branch II, where $\mathcal{K}^{abcd}$ is invertible and the degeneracy is encoded in the field-space matrix $\mathcal{S}_{ij}$.

For a single scalar field, this matrix reduces to a single coefficient. For $\mathcal{N}\ge2$, however, $\mathcal{S}_{ij}$ is an $\mathcal{N}\times\mathcal{N}$ matrix, and its rank contains information that has no counterpart in the single-field case. In particular, $\det\mathcal{S}=0$ only ensures that at least one combination of higher-derivative scalar modes is constrained. Removing all such modes requires enough independent primary constraints, and therefore requires a stronger condition on the rank of $\mathcal{S}_{ij}$. This point will be made precise in the next section.
\section{Complete degeneracy in the metric-invertible branch}
\label{Sec: Complete degeneracy in the metric-invertible branch}
We now implement the Branch II degeneracy of section \ref{Sec: Primary degeneracy conditions}, where $\mathcal{K}^{abcd}$ is invertible, Eq. \eqref{Eq: Branch II degeneracy condition I}, and the primary degeneracy is governed by the effective scalar matrix $\mathcal{S}_{ij}$ of Eq. \eqref{Eq: Branch II Schur complement}. The stronger condition anticipated there is complete degeneracy of the scalar sector, $\mathcal{S}_{ij}=0$. Because $\mathcal{S}_{ij}$ contains the inverse of $\mathcal{K}^{abcd}$, we represent that inverse through an auxiliary tensor and then reduce the condition to a set of conditions on the coefficient functions of the action.
\subsection{Auxiliary tensor for the inverse metric block}
\label{Subsec: Auxiliary tensor for the inverse metric block}
The Schur complement in Branch II contains the inverse of $\mathcal{K}^{abcd}$ as a map on the space of symmetric spatial tensors. Instead of writing this inverse explicitly, we introduce, for each field-space index $i$, a symmetric spatial tensor $\mathcal{W}_{iab}$ defined by
\begin{align}
\label{Eq: Auxiliary tensor definition}
\mathcal{K}^{abcd}\mathcal{W}_{icd}=\mathcal{B}_{i}{}^{ab}.
\end{align}

Since $\mathcal{K}^{abcd}$ is invertible in this branch, this equation uniquely determines $\mathcal{W}_{iab}$ as a tensor. Equivalently,
\begin{align}
\label{Eq: Auxiliary tensor inverse representation}
\mathcal{W}_{iab}=\left(\mathcal{K}^{-1}\right)_{abcd}\mathcal{B}_{i}{}^{cd}.
\end{align}

This auxiliary tensor is introduced only to represent the action of $\left(\mathcal{K}^{-1}\right)_{abcd}$ without computing the inverse metric block explicitly.

Because $\mathcal{K}^{abcd}$ and $\mathcal{B}_{i}{}^{ab}$ are built only from $h_{ab}$ and $\hat{\mathcal{A}}^{i}{}_{a}$, the map $\mathcal{K}^{abcd}$ preserves the span generated by $h_{ab}$ and $\hat{\mathcal{A}}^{j}{}_{a}\hat{\mathcal{A}}^{k}{}_{b}$. Therefore $\mathcal{W}_{iab}$ lies in the same span and admits, with no loss of generality, the expansion
\begin{align}
\label{Eq: Auxiliary tensor decomposition}
\mathcal{W}_{iab}=\alpha_{i}h_{ab}+\beta_{ijk}\hat{\mathcal{A}}^{j}{}_{a}\hat{\mathcal{A}}^{k}{}_{b},\qquad
\beta_{ijk}=\beta_{ikj}.
\end{align}

Here $\alpha_{i}$ and $\beta_{ijk}$ are coefficient functions, not new dynamical variables. For each fixed leading index $i$, the coefficient vector contains one scalar coefficient $\alpha_{i}$ and $\mathcal{N}(\mathcal{N}+1)/2$ symmetric-pair coefficients $\beta_{ijk}$. Thus the number of coefficient functions to be solved for each $i$ is
\begin{align}
\label{Eq: Number of auxiliary coefficients per field}
1+\frac{\mathcal{N}(\mathcal{N}+1)}{2}.
\end{align}

If the generating set is overcomplete, which can occur for $\mathcal{N}\ge3$ or on special field configurations, the coefficient functions need not be unique, but the tensor $\mathcal{W}_{iab}$ and the Schur complement constructed from it are unique.
\subsection{Coefficient system}
\label{Subsec: Coefficient system}
Acting with $\mathcal{K}^{abcd}$ on the decomposition \eqref{Eq: Auxiliary tensor decomposition} and expanding the result in the same basis gives
\begin{align}
\label{Eq: LHS decomposition}
\mathcal{K}^{abcd}\mathcal{W}_{icd}=\left(\mathsf{A}\alpha_{i}+\mathsf{B}^{jk}\beta_{ijk}\right)h^{ab}+\left(\mathsf{C}_{jk}\alpha_{i}+\mathsf{D}_{jk}{}^{lm}\beta_{ilm}\right)\hat{\mathcal{A}}^{ja}\hat{\mathcal{A}}^{kb},
\end{align}
where the coefficients $\mathsf{A}$, $\mathsf{B}^{ij}$, $\mathsf{C}_{ij}$ and $\mathsf{D}_{ij}{}^{kl}$ are collected in appendix \ref{AppSec: Coefficient system for the metric-invertible branch}.

Similarly, the mixing coefficient $\mathcal{B}_{i}{}^{ab}$ is decomposed in the same basis as
\begin{align}
\label{Eq: RHS decomposition}
\mathcal{B}_{i}{}^{ab}=\mathcal{Z}_{i}h^{ab}+\mathcal{Z}_{ijk}\hat{\mathcal{A}}^{ja}\hat{\mathcal{A}}^{kb},\qquad
\mathcal{Z}_{ijk}=\mathcal{Z}_{ikj}.
\end{align}

The explicit expressions for $\mathcal{Z}_{i}$ and $\mathcal{Z}_{ijk}$ are also given in appendix \ref{AppSec: Coefficient system for the metric-invertible branch}.

Matching Eqs. \eqref{Eq: LHS decomposition} and \eqref{Eq: RHS decomposition}, the defining equation \eqref{Eq: Auxiliary tensor definition} is equivalent to the coefficient system
\begin{align}
\label{Eq: Auxiliary tensor solution I}
&\mathsf{A}\alpha_{i}+\mathsf{B}^{jk}\beta_{ijk}=\mathcal{Z}_{i},\nonumber\\&
\mathsf{C}_{jk}\alpha_{i}+\mathsf{D}_{jk}{}^{lm}\beta_{ilm}=\mathcal{Z}_{ijk}.
\end{align}

For each fixed leading index $i$, this is a linear system on the coefficient space spanned by the scalar slot and the symmetric pair $(jk)$. In matrix notation,
\begin{align}
\label{Eq: Auxiliary tensor solution II}
\mathbb{M}
\begin{pmatrix}
\alpha_{i}\\
\beta_{ilm}
\end{pmatrix}
=
\begin{pmatrix}
\mathcal{Z}_{i}\\
\mathcal{Z}_{ijk}
\end{pmatrix},\qquad
\mathbb{M}=
\begin{pmatrix}
\mathsf{A}&\mathsf{B}^{lm}\\
\mathsf{C}_{jk}&\mathsf{D}_{jk}{}^{lm}
\end{pmatrix}.
\end{align}

The matrix $\mathbb{M}$ acts on the coefficient space of dimension $1+\mathcal{N}(\mathcal{N}+1)/2$, which may be larger than the dimension of the symmetric-tensor space represented by this basis. The leading index $i$ labels the source vector on the right-hand side and is not an index of $\mathbb{M}$. Solving Eq. \eqref{Eq: Auxiliary tensor solution II} determines the tensor $\mathcal{W}_{iab}$. If the expansion coefficients are not unique, they differ only by directions that leave $\mathcal{W}_{iab}$ unchanged, and therefore give the same Schur complement $\mathcal{S}_{ij}$.
\subsection{Effective scalar degeneracy matrix}
\label{Subsec: Effective scalar degeneracy matrix}
Using $\mathcal{W}_{iab}$, the Schur complement can be written as
\begin{align}
\label{Eq: Schur complement I}
\mathcal{S}_{ij}=\mathcal{G}_{ij}-\mathcal{B}_{i}{}^{ab}\mathcal{W}_{jab}.
\end{align}

Although the right-hand side is not manifestly symmetric in $i,j$, it is symmetric. Indeed, using Eq. \eqref{Eq: Auxiliary tensor inverse representation}, the second term is
\begin{align}
\label{Eq: Symmetry of Schur complement}
\mathcal{B}_{i}{}^{ab}\mathcal{W}_{jab}=\mathcal{B}_{i}{}^{ab}\left(\mathcal{K}^{-1}\right)_{abcd}\mathcal{B}_{j}{}^{cd},
\end{align}
which is symmetric under $i\leftrightarrow j$ because $\left(\mathcal{K}^{-1}\right)_{abcd}$ inherits the pair-exchange symmetry of $\mathcal{K}^{abcd}$.

Substituting Eq. \eqref{Eq: Auxiliary tensor decomposition} into Eq. \eqref{Eq: Schur complement I} gives
\begin{align}
\label{Eq: Schur complement II}
\mathcal{S}_{ij}=\mathcal{G}_{ij}-\mathcal{B}_{i}{}^{ab}\left(\alpha_{j}h_{ab}+\beta_{jkl}\hat{\mathcal{A}}^{k}{}_{a}\hat{\mathcal{A}}^{l}{}_{b}\right).
\end{align}

The required contractions $\mathcal{B}_{i}{}^{ab}h_{ab}$ and $\mathcal{B}_{i}{}^{ab}\hat{\mathcal{A}}^{k}{}_{a}\hat{\mathcal{A}}^{l}{}_{b}$, together with the expanded form of $\mathcal{S}_{ij}$, are given in appendix \ref{AppSec: Coefficient system for the metric-invertible branch}.
\subsection{Complete degeneracy condition}
\label{Subsec: Complete degeneracy condition}
In Branch II, the number of primary constraints is $\dim\ker\mathcal{S}$, Eq. \eqref{Eq: Number of primary constraints in branch II}. Therefore the condition $\det\mathcal{S}=0$ only guarantees at least one primary constraint. To remove all $\mathcal{N}$ higher-derivative scalar directions at the primary level, one must impose complete degeneracy of the effective scalar matrix,
\begin{align}
\label{Eq: Degeneracy condition}
\mathcal{S}_{ij}=0.
\end{align}

This is a system of $\mathcal{N}(\mathcal{N}+1)/2$ independent conditions on the coefficient functions of the action, since $\mathcal{S}_{ij}=\mathcal{S}_{ji}$. When these conditions hold, $\mathcal{S}_{ij}$ vanishes identically, its kernel is the full field space, and the velocity degeneracy produces $\mathcal{N}$ independent primary constraints, before imposing their preservation in time. The secondary constraint analysis will be carried out in the following sections. For $\mathcal{N}=1$, the matrix condition \eqref{Eq: Degeneracy condition} reduces to the single-field condition discussed in appendix \ref{AppSec: Single-field limit}.
\section{Hamiltonian analysis}
\label{Sec: Hamiltonian analysis}
We now perform the Hamiltonian analysis of the completely degenerate theory in the metric-invertible branch. The aim is to show how the primary degeneracy condition $\mathcal{S}_{ij}=0$ leads to primary constraints, how the preservation of these constraints can generate the required secondary constraints, and how the resulting constraint structure removes the Ostrogradsky modes associated with the variables $\mathcal{A}_{*}^{i}$. Related nonlinear degree-of-freedom analyses with dynamical gravity have been carried out for a single scalar field. These include the fully nonlinear Hamiltonian analysis of a beyond-Horndeski scalar-tensor theory in \cite{Lin:2014jga}, which establishes three physical degrees of freedom for the case considered there, and the analysis of generalized Galileons with a dynamical metric in \cite{Deffayet:2015qwa}, which includes a full Hamiltonian treatment of a particular model. We use the standard terminology of constrained Hamiltonian systems \cite{Dirac:1964,Henneaux:1992ig}. In particular, weak equality, denoted by $\approx$, means equality on the constraint surface; Poisson brackets must be computed before imposing weakly vanishing constraints.
\subsection{Canonical momenta}
\label{Subsec: Canonical momenta}
We start from the ADM Lagrangian density obtained in section \ref{Sec: ADM decomposition of the action}. We keep $\mathcal{U}$ unredefined, so that the contribution generated by integrating the acceleration term by parts appears explicitly as $-D_{a}\mathcal{E}^{a}$. Thus
\begin{align}
\label{Eq: Hamiltonian Lagrangian density}
\mathcal{L}&=\sqrt{h}N\Big[
F{}^{(3)}\!R
-D_{a}\mathcal{E}^{a}
+\mathcal{U}
+\mathcal{C}_{i}\mathcal{V}_{*}^{i}
+\mathcal{M}^{ab}K_{ab}
+\mathcal{G}_{ij}\mathcal{V}_{*}^{i}\mathcal{V}_{*}^{j}
+2\mathcal{B}_{i}{}^{ab}\mathcal{V}_{*}^{i}K_{ab}
+\mathcal{K}^{abcd}K_{ab}K_{cd}\nonumber\\&
+\lambda_{*i}(\mathcal{A}_{*}^{i}-\phi_{*}^{i})
+\hat{\lambda}_{i}{}^{a}(D_{a}\phi^{i}-\hat{\mathcal{A}}^{i}{}_{a})
\Big].
\end{align}

The extrinsic curvature is related to the time derivative of the spatial metric by
\begin{align}
\label{Eq: Hamiltonian extrinsic curvature definition}
K_{ab}=\frac{1}{2N}\left(\dot{h}_{ab}-D_{a}N_{b}-D_{b}N_{a}\right).
\end{align}

Therefore the momentum conjugate to $h_{ab}$ is
\begin{align}
\label{Eq: Momentum conjugate to h}
\pi^{ab}=\frac{\partial\mathcal{L}}{\partial\dot{h}_{ab}}=\sqrt{h}\left(\frac{1}{2}\mathcal{M}^{ab}+\mathcal{B}_{i}{}^{ab}\mathcal{V}_{*}^{i}+\mathcal{K}^{abcd}K_{cd}\right).
\end{align}

The velocity of $\mathcal{A}_{*}^{i}$ is contained in $\mathcal{V}_{*}^{i}$. Since
\begin{align}
\label{Eq: Normal derivative of Astar I}
n^{\mu}\nabla_{\mu}\mathcal{A}_{*}^{i}=\frac{1}{N}\left(\dot{\mathcal{A}}_{*}^{i}-N^{a}D_{a}\mathcal{A}_{*}^{i}\right),
\end{align}
and, using $\mathcal{A}_{*}^{i}=n^{\mu}\mathcal{A}^{i}{}_{\mu}$ together with the convention \eqref{Eq: Extrinsic curvature convention} and $a_{a}=D_{a}\ln{N}$,
\begin{align}
\label{Eq: Normal derivative of Astar II}
n^{\mu}\nabla_{\mu}\mathcal{A}_{*}^{i}=\mathcal{V}_{*}^{i}+\frac{1}{N}\hat{\mathcal{A}}^{ia}D_{a}N,
\end{align}
we have
\begin{align}
\label{Eq: Velocity variable Vstar}
\mathcal{V}_{*}^{i}=\frac{1}{N}\left(\dot{\mathcal{A}}_{*}^{i}-N^{a}D_{a}\mathcal{A}_{*}^{i}-\hat{\mathcal{A}}^{ia}D_{a}N\right).
\end{align}

The momentum conjugate to $\mathcal{A}_{*}^{i}$ is therefore
\begin{align}
\label{Eq: Momentum conjugate to Astar}
p_{*i}=\frac{\partial\mathcal{L}}{\partial\dot{\mathcal{A}}_{*}^{i}}=\sqrt{h}\left(\mathcal{C}_{i}+2\mathcal{G}_{ij}\mathcal{V}_{*}^{j}+2\mathcal{B}_{i}{}^{ab}K_{ab}\right).
\end{align}

For the scalar fields,
\begin{align}
\label{Eq: Velocity of scalar field}
\phi_{*}^{i}=n^{\mu}\nabla_{\mu}\phi^{i}=\frac{1}{N}\left(\dot{\phi}^{i}-N^{a}D_{a}\phi^{i}\right).
\end{align}

With the multiplier convention used in Eq. \eqref{Eq: Hamiltonian Lagrangian density}, the momentum conjugate to $\phi^{i}$ is
\begin{align}
\label{Eq: Momentum conjugate to phi}
p_{i}=\frac{\partial\mathcal{L}}{\partial\dot{\phi}^{i}}=-\sqrt{h}\lambda_{*i}.
\end{align}

The spatial auxiliary field $\hat{\mathcal{A}}^{i}{}_{a}$ has no time derivative in the Lagrangian, so that $\partial\mathcal{L}/\partial\dot{\hat{\mathcal{A}}}^{i}{}_{a}=0$ identically and its conjugate momentum is a primary constraint,
\begin{align}
\label{Eq: Momentum conjugate to Ahat}
\hat{p}_{i}{}^{a}\approx0.
\end{align}

Similarly, the lapse and shift are configuration variables of the ADM decomposition, but their time derivatives do not appear in the Lagrangian. Their conjugate momenta are therefore primary constraints,
\begin{align}
\label{Eq: Lapse and shift momenta}
p_{N}\approx0,\qquad
p_{a}\approx0.
\end{align}

The canonical pairs are
\begin{align}
\label{Eq: Hamiltonian canonical pairs}
(h_{ab},\pi^{ab}),\qquad
(\mathcal{A}_{*}^{i},p_{*i}),\qquad
(\phi^{i},p_{i}),\qquad
(\hat{\mathcal{A}}^{i}{}_{a},\hat{p}_{i}{}^{a}),\qquad
(N,p_{N}),\qquad
(N^{a},p_{a}).
\end{align}

These pairs fix the phase-space structure used in the constraint analysis that follows. The non-vanishing fundamental Poisson brackets are
\begin{align}
\label{Eq: Fundamental Poisson brackets fields}
&\{h_{ab}(x),\pi^{cd}(y)\}=\delta^{c}{}_{(a}\delta^{d}{}_{b)}\delta^{(3)}(x-y),\nonumber\\&
\{\mathcal{A}_{*}^{i}(x),p_{*j}(y)\}=\delta^{i}{}_{j}\delta^{(3)}(x-y),\nonumber\\&
\{\phi^{i}(x),p_{j}(y)\}=\delta^{i}{}_{j}\delta^{(3)}(x-y),\nonumber\\&
\{\hat{\mathcal{A}}^{i}{}_{a}(x),\hat{p}_{j}{}^{b}(y)\}=\delta^{i}{}_{j}\delta_{a}{}^{b}\delta^{(3)}(x-y),
\end{align}
together with those of the lapse and shift, listed separately because their momenta are the primary constraints associated with the ADM gauge variables,
\begin{align}
\label{Eq: Fundamental Poisson brackets lapse shift}
&\{N(x),p_{N}(y)\}=\delta^{(3)}(x-y),\nonumber\\&
\{N^{a}(x),p_{b}(y)\}=\delta^{a}{}_{b}\delta^{(3)}(x-y).
\end{align}
\subsection{Primary degeneracy constraints}
\label{Subsec: Primary degeneracy constraints}
We now impose the complete degeneracy condition $\mathcal{S}_{ij}=0$ of section \ref{Sec: Complete degeneracy in the metric-invertible branch} at the level of the canonical momenta, where $\mathcal{W}_{iab}$ is the auxiliary tensor defined in Eq. \eqref{Eq: Auxiliary tensor definition}. From Eqs. \eqref{Eq: Momentum conjugate to h} and \eqref{Eq: Momentum conjugate to Astar}, consider the combination $p_{*i}-2\mathcal{W}_{iab}\pi^{ab}$. The terms proportional to $K_{ab}$ cancel by Eq. \eqref{Eq: Auxiliary tensor definition}, and the terms proportional to $\mathcal{V}_{*}^{i}$ cancel by the degeneracy condition $\mathcal{S}_{ij}=0$, Eq. \eqref{Eq: Degeneracy condition}. Therefore
\begin{align}
\label{Eq: Primary degeneracy constraint derivation}
p_{*i}-2\mathcal{W}_{iab}\pi^{ab}=\sqrt{h}\left(\mathcal{C}_{i}-\mathcal{W}_{iab}\mathcal{M}^{ab}\right).
\end{align}

Defining
\begin{align}
\label{Eq: Hamiltonian Delta definition}
\Delta_{i}=\mathcal{W}_{iab}\mathcal{M}^{ab}-\mathcal{C}_{i},
\end{align}
we obtain the primary degeneracy constraints
\begin{align}
\label{Eq: Primary degeneracy constraints}
\Psi_{i}=p_{*i}-2\mathcal{W}_{iab}\pi^{ab}+\sqrt{h}\Delta_{i}\approx0.
\end{align}

These $\mathcal{N}$ primary constraints are the primary-level consequence of complete degeneracy. They remove the possibility of solving all scalar velocities $\mathcal{V}_{*}^{i}$ in terms of canonical momenta. By themselves, however, they do not yet guarantee that the Ostrogradsky modes are removed; this is decided by their preservation in time, which we analyze in the remainder of this section.

For the constraint analysis it is useful to separate $\Delta_{i}$ into a part free of spatial derivatives and the parts carrying spatial derivatives of the scalar and auxiliary fields,
\begin{align}
\label{Eq: Hamiltonian Delta decomposition}
\Delta_{i}=\Lambda_{i}+\mathcal{J}_{ij}{}^{a}D_{a}\mathcal{A}_{*}^{j}+\mathcal{I}_{ij}{}^{ab}S^{j}{}_{ab}.
\end{align}

Here $\Lambda_{i}$ is the part of $\Delta_{i}$ that contains no $D_{a}\mathcal{A}_{*}^{j}$ or $S^{j}{}_{ab}$, $\mathcal{J}_{ij}{}^{a}$ is the coefficient of $D_{a}\mathcal{A}_{*}^{j}$, and $\mathcal{I}_{ij}{}^{ab}$ is the coefficient of $S^{j}{}_{ab}$, symmetric in its spatial indices. The explicit expressions are collected in appendix \ref{AppSec: Explicit coefficients in the decomposition of Delta}.
\subsection{Canonical and total Hamiltonians}
\label{Subsec: Canonical and total Hamiltonians}
The canonical Hamiltonian is obtained from the Legendre transform
\begin{align}
\label{Eq: Canonical Hamiltonian definition}
H_{\rm{can}}=\int d^{3}x\left(\dot{h}_{ab}\pi^{ab}+\dot{\mathcal{A}}_{*}^{i}p_{*i}+\dot{\phi}^{i}p_{i}+\dot{\hat{\mathcal{A}}}^{i}{}_{a}\hat{p}_{i}{}^{a}-\mathcal{L}\right).
\end{align}

Using the velocity relations \eqref{Eq: Hamiltonian extrinsic curvature definition}, \eqref{Eq: Velocity variable Vstar} and \eqref{Eq: Velocity of scalar field} together with the momentum definitions \eqref{Eq: Momentum conjugate to phi} and \eqref{Eq: Momentum conjugate to Ahat}, and integrating by parts the terms containing derivatives of $N$ and $N^{a}$, the canonical Hamiltonian can be written as
\begin{align}
\label{Eq: Canonical Hamiltonian}
H_{\rm{can}}=\int d^{3}x\left(N\mathcal{H}_{0}+N^{a}\mathcal{H}_{a}+\ell_{i}{}^{a}\chi^{i}{}_{a}\right),
\end{align}
where
\begin{align}
\label{Eq: Auxiliary constraint chi}
\chi^{i}{}_{a}=\hat{\mathcal{A}}^{i}{}_{a}-D_{a}\phi^{i},
\end{align}
and
\begin{align}
\label{Eq: Auxiliary multiplier ell}
\ell_{i}{}^{a}=\sqrt{h}N\hat{\lambda}_{i}{}^{a}-p_{i}N^{a}.
\end{align}

The Hamiltonian density is
\begin{align}
\label{Eq: Hamiltonian constraint density}
\mathcal{H}_{0}&=p_{i}\mathcal{A}_{*}^{i}+\frac{1}{\sqrt{h}}\left(\pi^{ab}-\frac{1}{2}\sqrt{h}\mathcal{M}^{ab}\right)\left(\mathcal{K}^{-1}\right)_{abcd}\left(\pi^{cd}-\frac{1}{2}\sqrt{h}\mathcal{M}^{cd}\right)\nonumber\\&-\sqrt{h}\left(F{}^{(3)}\!R+\mathcal{U}-D_{a}\mathcal{E}^{a}\right)-D_{a}\left(p_{*i}\hat{\mathcal{A}}^{ia}\right),
\end{align}
and the momentum density is
\begin{align}
\label{Eq: Momentum constraint density}
\mathcal{H}_{a}=p_{i}\hat{\mathcal{A}}^{i}{}_{a}+p_{*i}D_{a}\mathcal{A}_{*}^{i}-2D_{b}\pi_{a}{}^{b}.
\end{align}

The expression \eqref{Eq: Momentum constraint density} is the form relevant on the auxiliary constraint surface $\chi^{i}{}_{a}\approx0$ and $\hat{p}_{i}{}^{a}\approx0$. The full spatial-diffeomorphism generator differs from Eq. \eqref{Eq: Momentum constraint density} by terms proportional to $\chi^{i}{}_{a}$ and $\hat{p}_{i}{}^{a}$, which vanish on the auxiliary constraint surface and do not affect the degeneracy analysis.

The term $\ell_{i}{}^{a}\chi^{i}{}_{a}$ in Eq. \eqref{Eq: Canonical Hamiltonian} shows that $\chi^{i}{}_{a}\approx0$ is an auxiliary primary constraint. In the total Hamiltonian, every primary constraint is included with an arbitrary multiplier. The specific coefficient $\ell_{i}{}^{a}$ is therefore absorbed into a new arbitrary multiplier. The total Hamiltonian is
\begin{align}
\label{Eq: Total Hamiltonian}
H_{\rm{T}}=\int d^{3}x\Big(
N\mathcal{H}_{0}
+N^{a}\mathcal{H}_{a}
+u^{i}\Psi_{i}
+v_{N}p_{N}
+v^{a}p_{a}
+\mu^{i}{}_{a}\hat{p}_{i}{}^{a}
+\nu_{i}{}^{a}\chi^{i}{}_{a}
\Big).
\end{align}

Here $u^{i}$, $v_{N}$, $v^{a}$, $\mu^{i}{}_{a}$ and $\nu_{i}{}^{a}$ are Lagrange multipliers enforcing the primary constraints $\Psi_{i}\approx0$, $p_{N}\approx0$, $p_{a}\approx0$, $\hat{p}_{i}{}^{a}\approx0$ and $\chi^{i}{}_{a}\approx0$, respectively.
\subsection{Auxiliary constraints and reduced phase space}
\label{Subsec: Auxiliary constraints and reduced phase space}
Preservation of the lapse and shift momenta gives
\begin{align}
\label{Eq: Preservation of lapse momentum}
\dot{p}_{N}=\{p_{N},H_{\rm{T}}\}=-\frac{\delta H_{\rm{T}}}{\delta N}=-\mathcal{H}_{0}\approx0,
\end{align}
and
\begin{align}
\label{Eq: Preservation of shift momentum}
\dot{p}_{a}=\{p_{a},H_{\rm{T}}\}=-\frac{\delta H_{\rm{T}}}{\delta N^{a}}=-\mathcal{H}_{a}\approx0.
\end{align}

Thus we obtain the secondary constraints
\begin{align}
\label{Eq: ADM secondary constraints}
\mathcal{H}_{0}\approx0,\qquad
\mathcal{H}_{a}\approx0.
\end{align}

They are the Hamiltonian and momentum constraints associated with spacetime diffeomorphism invariance.

We next consider the auxiliary constraints
\begin{align}
\label{Eq: Auxiliary constraints}
\chi^{i}{}_{a}\approx0,\qquad
\hat{p}_{i}{}^{a}\approx0.
\end{align}

Their non-vanishing Poisson bracket is
\begin{align}
\label{Eq: Auxiliary second class bracket}
\{\chi^{i}{}_{a}(x),\hat{p}_{j}{}^{b}(y)\}=\delta^{i}{}_{j}\delta_{a}{}^{b}\delta^{(3)}(x-y),
\end{align}
so this pair is second class. This means that the matrix of their mutual Poisson brackets is invertible, and their preservation fixes the associated multipliers rather than generating new independent constraints.

Indeed,
\begin{align}
\label{Eq: Preservation of chi I}
\dot{\chi}^{i}{}_{a}=\{\chi^{i}{}_{a},H_{\rm{T}}\}=\{\hat{\mathcal{A}}^{i}{}_{a},H_{\rm{T}}\}-D_{a}\{\phi^{i},H_{\rm{T}}\}.
\end{align}

Using
\begin{align}
\label{Eq: Bracket Ahat total Hamiltonian}
\{\hat{\mathcal{A}}^{i}{}_{a},H_{\rm{T}}\}=\mu^{i}{}_{a},
\end{align}
and
\begin{align}
\label{Eq: Bracket phi total Hamiltonian}
\{\phi^{i},H_{\rm{T}}\}=N\mathcal{A}_{*}^{i}+N^{a}\hat{\mathcal{A}}^{i}{}_{a},
\end{align}
we find
\begin{align}
\label{Eq: Preservation of chi II}
\dot{\chi}^{i}{}_{a}=\mu^{i}{}_{a}-D_{a}\left(N\mathcal{A}_{*}^{i}+N^{b}\hat{\mathcal{A}}^{i}{}_{b}\right)\approx0.
\end{align}

Therefore
\begin{align}
\label{Eq: Auxiliary multiplier mu}
\mu^{i}{}_{a}\approx D_{a}\left(N\mathcal{A}_{*}^{i}+N^{b}\hat{\mathcal{A}}^{i}{}_{b}\right).
\end{align}

Similarly, preservation of $\hat{p}_{i}{}^{a}\approx0$ gives
\begin{align}
\label{Eq: Preservation of phat I}
\dot{\hat{p}}_{i}{}^{a}=\{\hat{p}_{i}{}^{a},H_{\rm{T}}\}=-\frac{\delta H_{\rm{T}}}{\delta\hat{\mathcal{A}}^{i}{}_{a}}\approx0.
\end{align}

Since $H_{\rm{T}}$ depends on $\hat{\mathcal{A}}^{i}{}_{a}$ both algebraically and through $S^{i}{}_{ab}=D_{(a}\hat{\mathcal{A}}^{i}{}_{b)}$, this is a functional derivative. Isolating the multiplier $\nu_{i}{}^{a}$ gives
\begin{align}
\label{Eq: Auxiliary multiplier nu}
\nu_{i}{}^{a}\approx
-\frac{\delta H_{\rm{can}}}{\delta\hat{\mathcal{A}}^{i}{}_{a}}
-\frac{\delta}{\delta\hat{\mathcal{A}}^{i}{}_{a}}\int d^{3}xu^{j}\Psi_{j}.
\end{align}

Thus preservation of $\hat{p}_{i}{}^{a}\approx0$ fixes $\nu_{i}{}^{a}$ and does not produce a new constraint.

Since the auxiliary pair is second class, it may be eliminated by the corresponding Dirac bracket. Equivalently, after this reduction one can impose $\chi^{i}{}_{a}=0$ and $\hat{p}_{i}{}^{a}=0$ strongly. We denote the resulting bracket by $\{\ ,\ \}_{D,{\rm{aux}}}$ and the reduced canonical Hamiltonian by
\begin{align}
\label{Eq: Reduced canonical Hamiltonian}
H_{\rm{red}}=\int d^{3}x\left(N\mathcal{H}_{0}+N^{a}\mathcal{H}_{a}\right),
\end{align}
with the understanding that the auxiliary constraints have been imposed. The reduced total Hamiltonian is
\begin{align}
\label{Eq: Reduced total Hamiltonian}
H_{\rm{T}}^{\rm{red}}=H_{\rm{red}}+\int d^{3}x\left(u^{i}\Psi_{i}+v_{N}p_{N}+v^{a}p_{a}\right).
\end{align}
\subsection{Preservation of the primary degeneracy constraints}
\label{Subsec: Preservation of the primary degeneracy constraints}
We now impose preservation of the primary degeneracy constraints $\Psi_{i}\approx0$. Since $\Psi_{i}$ contains $D_{a}\mathcal{A}_{*}^{i}$ and $S^{i}{}_{ab}=D_{(a}\hat{\mathcal{A}}^{i}{}_{b)}$ through $\Delta_{i}$, the local bracket $\{\Psi_{i}(x),\Psi_{j}(y)\}$ contains derivatives of the delta distribution. It is therefore more transparent to work with smeared constraints,
\begin{align}
\label{Eq: Smeared primary constraint}
\Psi[f]=\int d^{3}xf^{i}\Psi_{i},
\end{align}
where $f^{i}$ is an arbitrary test function. The preservation condition is
\begin{align}
\label{Eq: Preservation of smeared primary constraint}
\dot{\Psi}[f]=\{\Psi[f],H_{\rm{T}}^{\rm{red}}\}_{D,{\rm{aux}}}\approx0.
\end{align}

By construction, $\Psi_{i}$ depends only on the canonical variables and has no explicit dependence on $N$ or $N^{a}$. Therefore
\begin{align}
\label{Eq: Psi lapse shift brackets vanish}
\{\Psi[f],p_{N}[v_{N}]\}_{D,{\rm{aux}}}=0,\qquad
\{\Psi[f],p_{a}[v^{a}]\}_{D,{\rm{aux}}}=0,
\end{align}
where
\begin{align}
\label{Eq: Smeared lapse shift momenta}
p_{N}[v_{N}]=\int d^{3}xv_{N}p_{N},\qquad
p_{a}[v^{a}]=\int d^{3}xv^{a}p_{a}.
\end{align}

Thus
\begin{align}
\label{Eq: Preservation of Psi reduced}
\dot{\Psi}[f]=\{\Psi[f],H_{\rm{red}}\}_{D,{\rm{aux}}}+\{\Psi[f],\Psi[u]\}_{D,{\rm{aux}}}\approx0.
\end{align}

The second term is the possible obstruction to obtaining secondary constraints. If it contains the arbitrary multiplier $u^{i}$, or derivatives of $u^{i}$, with non-vanishing coefficients, then Eq. \eqref{Eq: Preservation of Psi reduced} fixes $u^{i}$ instead of producing a new constraint. Therefore, to generate secondary constraints associated with $\Psi_{i}\approx0$, the primary-primary bracket must vanish weakly as a differential operator acting on the smearing functions.

To analyze this operator, we compute $\{\Psi[f],\Psi[g]\}_{D,{\rm{aux}}}$ for two arbitrary test functions $f^{i}$ and $g^{i}$. The preservation equation is recovered by setting $g^{i}=u^{i}$. With this convention, derivative terms in $g^{i}$ become derivative terms in the multiplier $u^{i}$. Computing these brackets and integrating by parts until at most one derivative acts on the smearing functions, the bracket can be organized as
\begin{align}
\label{Eq: Smeared primary primary structure}
\{\Psi[f],\Psi[g]\}_{D,{\rm{aux}}}=\int d^{3}xf^{i}g^{j}\mathfrak{C}_{ij}+\int d^{3}x\left(f^{i}D_{a}g^{j}-g^{i}D_{a}f^{j}\right)\mathfrak{D}_{ij}{}^{a}.
\end{align}

The coefficient $\mathfrak{D}_{ij}{}^{a}$ multiplies the derivative-smearing structure and would therefore multiply $D_{a}u^{j}$ in $\{\Psi[f],\Psi[u]\}_{D,{\rm{aux}}}$. The coefficient $\mathfrak{C}_{ij}$ multiplies the non-derivative smearing structure. We decompose it as
\begin{align}
\label{Eq: C decomposition}
\mathfrak{C}_{ij}=\mathfrak{M}_{ij}+\mathfrak{N}_{ij},
\end{align}
where $\mathfrak{M}_{ij}$ is the part proportional to the canonical momentum $\pi^{ab}$, while $\mathfrak{N}_{ij}$ is independent of $\pi^{ab}$. Because the Poisson bracket changes sign under exchange of the two smearing functions, the non-derivative coefficient is antisymmetric in field space, $\mathfrak{C}_{ij}=-\mathfrak{C}_{ji}$, and hence so are $\mathfrak{M}_{ij}$ and $\mathfrak{N}_{ij}$. Both therefore vanish identically for a single scalar field: they are intrinsically multi-field conditions with no single-field analogue.

That preservation of the primary constraints can impose conditions beyond primary degeneracy is not specific to the present setting. For coupled systems with several higher-derivative variables, it was shown in \cite{Motohashi:2016ftl,Klein:2016aiq} that degeneracy of the kinetic matrix is in general not sufficient and that further conditions are needed, and the corresponding field-theory analysis of \cite{Crisostomi:2017aim} organized the secondary conditions into symmetric and antisymmetric parts in the field-space indices, under assumptions that exclude gauge symmetries and a dynamical spin-2 field. The present subsection derives the primary-constraint consistency conditions in the corresponding generally covariant multi-scalar-tensor setting, with a dynamical metric and arbitrary $\mathcal{N}$, in which the scalar and metric velocities mix through the ADM Hessian and the gravitational constraint sector is present. In this setting the consistency coefficients are obtained explicitly in terms of the ADM quantities generated by the covariant action, including $\mathcal{W}_{iab}$, $\mathcal{J}_{ij}{}^{a}$, $\mathcal{I}_{ij}{}^{ab}$, and $\Delta_{i}$, as given below.

Before writing the explicit expressions, we define the algebraic derivative used below. The subscript ${\rm{alg}}$ denotes the derivative with respect to the independent spatial fields $h_{ab}$, $\mathcal{A}_{*}^{i}$ and $\hat{\mathcal{A}}^{i}{}_{a}$, acting on all explicit occurrences, including the dependence of the kinetic matrix on these fields through its ADM form \eqref{Eq: Decomposition of kinetic matrix}, in which $\hat{\mathcal{A}}^{ij}=h^{ab}\hat{\mathcal{A}}^{i}{}_{a}\hat{\mathcal{A}}^{j}{}_{b}$ carries the dependence on $h_{ab}$. It does not act on the explicit spatial derivatives $D_{a}\mathcal{A}_{*}^{i}$ and $S^{i}{}_{ab}$. These are ordinary partial derivatives of the coefficient functions, not functional derivatives.

The constraint $\Psi_{i}$ depends on $\hat{\mathcal{A}}^{i}{}_{a}$ not only algebraically but also through $S^{i}{}_{ab}=D_{(a}\hat{\mathcal{A}}^{i}{}_{b)}$, which enters $\Delta_{i}$ with coefficient $\mathcal{I}_{ij}{}^{ab}$ in Eq. \eqref{Eq: Hamiltonian Delta decomposition}. In the primary-primary bracket this dependence is carried by the variation of the spatial connection inside $S^{i}{}_{ab}$, which we collect in the tensor
\begin{align}
\label{Eq: R tensor definition}
\mathcal{R}_{i}{}^{abc}=-\frac{1}{2}\hat{\mathcal{A}}^{jc}\mathcal{I}_{ij}{}^{ab}+\hat{\mathcal{A}}^{j(a}\mathcal{I}_{ij}{}^{b)c}.
\end{align}

This tensor enters the derivative-smearing coefficient $\mathfrak{D}_{ij}{}^{a}$ and, through $D_{c}\mathcal{R}_{j}{}^{abc}$, the momentum-independent coefficient $\mathfrak{N}_{ij}$. In terms of it, the derivative-smearing coefficient is
\begin{align}
\label{Eq: Derivative smearing coefficient}
\mathfrak{D}_{ij}{}^{a}=\sqrt{h}\left(\mathcal{J}_{ji}{}^{a}+2\mathcal{W}_{ibc}\mathcal{R}_{j}{}^{bca}\right).
\end{align}

Collecting the terms linear in $\pi^{ab}$, the momentum-dependent coefficient is
\begin{align}
\mathfrak{M}_{ij}&=\bigg[2\left(\frac{\partial\mathcal{W}_{jab}}{\partial\mathcal{A}_{*}^{i}}-\frac{\partial\mathcal{W}_{iab}}{\partial\mathcal{A}_{*}^{j}}\right)_{\rm{alg}}+4\left(\left(\frac{\partial\mathcal{W}_{iab}}{\partial h_{cd}}\right)_{\rm{alg}}\mathcal{W}_{jcd}-\mathcal{W}_{icd}\left(\frac{\partial\mathcal{W}_{jab}}{\partial h_{cd}}\right)_{\rm{alg}}\right)\bigg]\pi^{ab}.\nonumber\\&
\label{Eq: Momentum dependent bracket coefficient}
\end{align}

Collecting the terms independent of $\pi^{ab}$, the momentum-independent coefficient is
\begin{align}
\label{Eq: Momentum independent bracket coefficient}
\mathfrak{N}_{ij}&=\sqrt{h}\bigg[\left(\frac{\partial\Delta_{i}}{\partial\mathcal{A}_{*}^{j}}-\frac{\partial\Delta_{j}}{\partial\mathcal{A}_{*}^{i}}\right)_{\rm{alg}}-2\left(\left(\frac{\partial\Delta_{i}}{\partial h_{ab}}\right)_{\rm{alg}}\mathcal{W}_{jab}-\mathcal{W}_{iab}\left(\frac{\partial\Delta_{j}}{\partial h_{ab}}\right)_{\rm{alg}}\right)\nonumber\\&+2\mathcal{W}_{[i|a|}{}^{a}\Delta_{j]}-2D_{a}\mathcal{J}_{[ij]}{}^{a}+4\mathcal{W}_{[i|ab|}D_{c}\mathcal{R}_{j]}{}^{abc}\bigg].
\end{align}

Here brackets on field-space indices denote weighted antisymmetrization, for example
\begin{align}
\label{Eq: Weighted antisymmetrization}
T_{[ij]}=\frac{1}{2}\left(T_{ij}-T_{ji}\right).
\end{align}

With the organization in Eq. \eqref{Eq: Smeared primary primary structure}, the terms associated with derivatives of the smearing functions are kept in $\mathfrak{D}_{ij}{}^{a}$ and are not moved into $\mathfrak{N}_{ij}$ by further integration by parts. Equivalently, the split between $\mathfrak{D}_{ij}{}^{a}$ and $\mathfrak{N}_{ij}$ is partly a convention for organizing the smeared bracket, while the full differential-operator condition is invariant. Appendix \ref{AppSec: Organization of the smeared primary-primary bracket} describes the assembly of this bracket and the origin of the individual terms in $\mathfrak{D}_{ij}{}^{a}$, $\mathfrak{M}_{ij}$, and $\mathfrak{N}_{ij}$.

The necessary Dirac requirement is weak vanishing of the primary-primary bracket on the already imposed constraint surface. Equivalently, all coefficients in Eq. \eqref{Eq: Smeared primary primary structure} must vanish weakly,
\begin{align}
\label{Eq: Weak primary primary conditions}
\mathfrak{D}_{ij}{}^{a}\approx0,\qquad
\mathfrak{M}_{ij}\approx0,\qquad
\mathfrak{N}_{ij}\approx0.
\end{align}

Weak vanishing means that these quantities may vanish only after using the constraints already obtained. For a theory-level classification, however, it is useful to impose a stronger sufficient requirement: the same coefficients are set to zero as identities on phase space,
\begin{align}
\label{Eq: Strong primary primary conditions}
\mathfrak{D}_{ij}{}^{a}=0,\qquad
\mathfrak{M}_{ij}=0,\qquad
\mathfrak{N}_{ij}=0.
\end{align}

These strong conditions are not logically necessary in the most general Dirac analysis, but they guarantee that the multiplier $u^{i}$ is not fixed by the preservation of $\Psi_{i}\approx0$.

Since $\sqrt{h}\neq0$ in the non-degenerate ADM branch, the strong version of $\mathfrak{D}_{ij}{}^{a}=0$ is equivalently the coefficient condition
\begin{align}
\label{Eq: Strong derivative smearing coefficient condition}
\mathcal{J}_{ji}{}^{a}+2\mathcal{W}_{ibc}\mathcal{R}_{j}{}^{bca}=0.
\end{align}

Since $\pi^{ab}$ is an independent canonical momentum, the strong version of $\mathfrak{M}_{ij}=0$ is equivalently the coefficient condition
\begin{align}
\label{Eq: Strong momentum coefficient condition}
\left(\frac{\partial\mathcal{W}_{jab}}{\partial\mathcal{A}_{*}^{i}}-\frac{\partial\mathcal{W}_{iab}}{\partial\mathcal{A}_{*}^{j}}\right)_{\rm{alg}}+2\left(\left(\frac{\partial\mathcal{W}_{iab}}{\partial h_{cd}}\right)_{\rm{alg}}\mathcal{W}_{jcd}-\mathcal{W}_{icd}\left(\frac{\partial\mathcal{W}_{jab}}{\partial h_{cd}}\right)_{\rm{alg}}\right)=0.
\end{align}

Similarly, the strong version of $\mathfrak{N}_{ij}=0$ is
\begin{align}
\label{Eq: Strong nonmomentum coefficient condition}
&\left(\frac{\partial\Delta_{i}}{\partial\mathcal{A}_{*}^{j}}-\frac{\partial\Delta_{j}}{\partial\mathcal{A}_{*}^{i}}\right)_{\rm{alg}}-2\left(\left(\frac{\partial\Delta_{i}}{\partial h_{ab}}\right)_{\rm{alg}}\mathcal{W}_{jab}-\mathcal{W}_{iab}\left(\frac{\partial\Delta_{j}}{\partial h_{ab}}\right)_{\rm{alg}}\right)\nonumber\\&+2\mathcal{W}_{[i|a|}{}^{a}\Delta_{j]}-2D_{a}\mathcal{J}_{[ij]}{}^{a}+4\mathcal{W}_{[i|ab|}D_{c}\mathcal{R}_{j]}{}^{abc}=0.
\end{align}

When the weak conditions \eqref{Eq: Weak primary primary conditions} hold, the primary-primary bracket vanishes weakly,
\begin{align}
\label{Eq: Psi Psi weak zero}
\{\Psi[f],\Psi[u]\}_{D,{\rm{aux}}}\approx0,
\end{align}
for arbitrary $f^{i}$ and arbitrary multipliers $u^{i}$. The preservation equation \eqref{Eq: Preservation of Psi reduced} then gives the secondary constraints
\begin{align}
\label{Eq: Secondary constraint smeared}
\Omega[f]\equiv\{\Psi[f],H_{\rm{red}}\}_{D,{\rm{aux}}}\approx0.
\end{align}

Equivalently, in local notation,
\begin{align}
\label{Eq: Secondary constraint local}
\Omega_{i}\equiv\{\Psi_{i},H_{\rm{red}}\}_{D,{\rm{aux}}}\approx0.
\end{align}

These are the secondary constraints associated with the primary degeneracy constraints $\Psi_{i}\approx0$.
\subsection{Secondary constraints and rank condition}
\label{Subsec: Secondary constraints and rank condition}
The final step in the scalar degeneracy sector is to preserve the secondary constraints. Introducing the smeared constraint
\begin{align}
\label{Eq: Smeared Omega}
\Omega[\eta]=\int d^{3}x\eta^{i}\Omega_{i},
\end{align}
we require
\begin{align}
\label{Eq: Preservation of Omega}
\dot{\Omega}[\eta]=\{\Omega[\eta],H_{\rm{T}}^{\rm{red}}\}_{D,{\rm{aux}}}\approx0.
\end{align}

At this stage the reduced total Hamiltonian contains the multiplier term $\Psi[u]$, so Eq. \eqref{Eq: Preservation of Omega} involves the bracket between $\Omega_{i}$ and $\Psi_{j}$ acting on the multipliers $u^{j}$. If this primary-secondary bracket is non-degenerate in the scalar sector, preservation of $\Omega_{i}\approx0$ fixes the multipliers $u^{i}$ rather than generating further constraints.

Since the constraints contain spatial derivatives, the bracket $\{\Omega_{i},\Psi_{j}\}_{D,{\rm{aux}}}$ should be understood as a matrix of differential operators acting on the smearing functions or on the multipliers $u^{j}$. The required condition is that this operator be invertible in the scalar sector, up to possible boundary zero modes or genuine gauge modes. Symbolically, we write
\begin{align}
\label{Eq: Rank condition}
\operatorname{rank}\left(\{\Omega_{i},\Psi_{j}\}_{D,{\rm{aux}}}\right)=\mathcal{N}.
\end{align}

Under this rank condition, the equations $\dot{\Omega}_{i}\approx0$ determine the $\mathcal{N}$ multipliers $u^{i}$ and no further scalar constraints are generated. Thus the pairs $\left(\Psi_{i},\Omega_{i}\right)$ are second class in the scalar degeneracy sector.
\subsection{Counting degrees of freedom}
\label{Subsec: Counting degrees of freedom}
We now count the physical degrees of freedom. We assume that the metric block $\mathcal{K}^{abcd}$ is invertible, complete primary degeneracy $\mathcal{S}_{ij}=0$ holds, the primary-primary bracket conditions \eqref{Eq: Weak primary primary conditions} hold, the rank condition \eqref{Eq: Rank condition} is satisfied, and the scalar second-class pair $(\Psi_{i},\Omega_{i})$ leaves the diffeomorphism constraints first class.

Before eliminating the auxiliary pair, the canonical pairs are those of Eq. \eqref{Eq: Hamiltonian canonical pairs}, with total phase-space dimension
\begin{align}
\label{Eq: Full phase space dimension}
12+2\mathcal{N}+2\mathcal{N}+6\mathcal{N}+2+6=20+10\mathcal{N}.
\end{align}

The auxiliary constraints $\chi^{i}{}_{a}\approx0$ and $\hat{p}_{i}{}^{a}\approx0$ are second class and remove $6\mathcal{N}$ phase-space dimensions. By the assumption above on the diffeomorphism sector, $p_{N}$, $\mathcal{H}_{0}$, $p_{a}$ and $\mathcal{H}_{a}$ remain first class on the reduced phase space, so the ADM constraints
\begin{align}
\label{Eq: ADM first class constraints}
p_{N}\approx0,\qquad
\mathcal{H}_{0}\approx0,\qquad
p_{a}\approx0,\qquad
\mathcal{H}_{a}\approx0,
\end{align}
are first class and remove $16$ phase-space dimensions. Finally, the $\mathcal{N}$ primary constraints $\Psi_{i}\approx0$ and the $\mathcal{N}$ secondary constraints $\Omega_{i}\approx0$ form $\mathcal{N}$ second-class pairs in the scalar degeneracy sector, and remove $2\mathcal{N}$ phase-space dimensions. Therefore the final physical phase-space dimension is
\begin{align}
\label{Eq: Final physical phase space dimension}
20+10\mathcal{N}-6\mathcal{N}-16-2\mathcal{N}=4+2\mathcal{N}.
\end{align}

The number of physical degrees of freedom is half of this number:
\begin{align}
\label{Eq: Final number of degrees of freedom}
N_{\rm{dof}}=\frac{1}{2}\left(4+2\mathcal{N}\right)=2+\mathcal{N}.
\end{align}

Thus the theory propagates the two tensor degrees of freedom of the metric and one scalar degree of freedom for each of the $\mathcal{N}$ scalar fields. The additional Ostrogradsky modes associated with the higher derivatives are removed by the second-class pairs $\left(\Psi_{i},\Omega_{i}\right)$.
\section{Consistency checks and special subclasses}
\label{Sec: Consistency checks and special subclasses}
In this section we apply the degeneracy and consistency conditions of section \ref{Sec: Hamiltonian analysis} to several special cases. The purpose is twofold. First, we check that the general construction reduces correctly in the single-field limit. Second, we examine two multi-field subclasses: a Horndeski-type quadratic subclass, for which most conditions simplify but a nontrivial closure condition remains, and a constructive ADM-level subclass for which the full set of strong conditions can be satisfied explicitly.
\subsection{Single-field case}
\label{Subsec: Single-field case}
We first consider the limit $\mathcal{N}=1$. The field-space indices then take a single value and are suppressed (so that $\mathcal{W}_{iab}\to\mathcal{W}_{ab}$, $\mathfrak{M}_{ij}\to\mathfrak{M}$, and so on), and the primary degeneracy condition reduces to a single scalar condition. The explicit reduction of the auxiliary tensor equation and of the primary degeneracy condition is given in appendix \ref{AppSec: Single-field limit}. In particular, the auxiliary tensor takes the single-field form of the expansion \eqref{Eq: Auxiliary tensor decomposition},
\begin{align}
\label{Eq: Single-field auxiliary tensor in special section}
\mathcal{W}_{ab}=\alpha h_{ab}+\beta\hat{\mathcal{A}}_{a}\hat{\mathcal{A}}_{b},
\end{align}
where the coefficients $\alpha$ and $\beta$ are determined by the single-field version of Eq. \eqref{Eq: Auxiliary tensor definition}. Substituting this solution into the Schur complement gives the single-field primary degeneracy condition displayed in appendix \ref{AppSec: Single-field limit}, which provides a check of the present construction against the degeneracy structure known for single-field quadratic higher-order scalar-tensor theories \cite{Langlois:2015skt}.

The primary-primary bracket conditions simplify in a special way. The coefficients $\mathfrak{M}_{ij}$ and $\mathfrak{N}_{ij}$ in Eqs. \eqref{Eq: Momentum dependent bracket coefficient} and \eqref{Eq: Momentum independent bracket coefficient}, which make up the non-derivative part $\mathfrak{C}_{ij}$ of the smeared bracket, are antisymmetric in the field-space indices $i,j$. With only one field they therefore vanish identically,
\begin{align}
\label{Eq: Single-field M and N vanish}
\mathfrak{M}=0,\qquad
\mathfrak{N}=0.
\end{align}

This argument does not by itself remove the derivative-smearing part, because even with a single constraint density the smeared bracket may have the form
\begin{align}
\label{Eq: Single-field derivative smearing possibility}
\{\Psi[f],\Psi[g]\}_{D,{\rm{aux}}}=\int d^{3}x\left(fD_{a}g-gD_{a}f\right)\mathfrak{D}^{a}.
\end{align}

Thus $\mathfrak{D}^{a}$ must be checked separately. From Eq. \eqref{Eq: Derivative smearing coefficient}, the single-field derivative-smearing coefficient is
\begin{align}
\label{Eq: Single-field derivative smearing coefficient}
\mathfrak{D}^{a}=\sqrt{h}\left(\mathcal{J}^{a}+2\mathcal{W}_{bc}\mathcal{R}^{bca}\right).
\end{align}

As shown in appendix \ref{AppSubsec: Reduction of the primary degeneracy condition}, the numerator of the single-field primary degeneracy condition decomposes, using the auxiliary tensor solution \eqref{Eq: Single-field auxiliary tensor solution}, into three independent coefficient equations, Eq. \eqref{Eq: Single-field degeneracy conditions}. Substituting these conditions, together with the single-field reductions of $\mathcal{J}^{a}$ and $\mathcal{I}^{ab}$ from appendix \ref{AppSec: Explicit coefficients in the decomposition of Delta}, into the parenthesis in Eq. \eqref{Eq: Single-field derivative smearing coefficient}, one finds that it vanishes; this substitution is carried out in appendix \ref{AppSubsec: Primary-primary consistency in the single-field limit}. Thus
\begin{align}
\label{Eq: Single-field D vanishes}
\mathfrak{D}^{a}=0.
\end{align}

Therefore, in the single-field case, the primary-primary bracket imposes no condition beyond the primary degeneracy condition: the antisymmetric conditions $\mathfrak{M}=\mathfrak{N}=0$ are trivial, and $\mathfrak{D}^{a}=0$ follows once the degeneracy condition holds. The preservation of the primary constraint then generates the secondary constraint which removes the Ostrogradsky mode, provided the usual primary-secondary bracket is non-degenerate. Thus, in the single-field limit, primary degeneracy is sufficient to pass the primary-primary consistency test.
\subsection{A multi-field quadratic Horndeski-type subclass}
\label{Subsec: A multi-field quadratic Horndeski-type subclass}
We next consider the multi-field quadratic Horndeski-type choice
\begin{align}
\label{Eq: Multi-field Horndeski coefficients}
A^{(1)}_{ij}=-A^{(2)}_{ij}=2F_{X^{ij}},\qquad
B_{ijk}=A^{(3)}_{ijkl}=A^{(4)}_{ijkl}=A^{(5)}_{ijklmn}=0.
\end{align}

The action then becomes
\begin{align}
\label{Eq: Multi-field Horndeski action}
S&=\int d^{4}x\sqrt{-g}\Big[F(\phi^{i},X^{ij})R+P(\phi^{i},X^{ij})+Q_{i}(\phi^{i},X^{ij})\phi^{i}{}_{\mu}{}^{\mu}\nonumber\\&+2F_{X^{ij}}(\phi^{i},X^{ij})\left(L_{1}^{ij}-L_{2}^{ij}\right)
\Big].
\end{align}

Substituting Eq. \eqref{Eq: Multi-field Horndeski coefficients} into the ADM coefficients gives
\begin{align}
\label{Eq: Multi-field Horndeski B and G}
\mathcal{G}_{ij}=0,\qquad
\mathcal{B}_{i}{}^{ab}=0.
\end{align}

Since $\mathcal{K}^{abcd}$ is invertible in the branch considered here, Eq. \eqref{Eq: Auxiliary tensor definition} then implies
\begin{align}
\label{Eq: Multi-field Horndeski W}
\mathcal{W}_{iab}=0.
\end{align}

This will be used below in the analysis of the primary-primary bracket. The primary degeneracy condition is automatically satisfied because $\mathcal{B}_{i}{}^{ab}$ and $\mathcal{G}_{ij}$ vanish, independently of $\mathcal{W}_{iab}$:
\begin{align}
\label{Eq: Multi-field Horndeski primary degeneracy}
\mathcal{S}_{ij}=\mathcal{G}_{ij}-\mathcal{B}_{i}{}^{ab}\mathcal{W}_{jab}=0.
\end{align}

The decomposition of $\Delta_{i}$ also simplifies. From appendix \ref{AppSec: Explicit coefficients in the decomposition of Delta}, one finds
\begin{align}
\label{Eq: Multi-field Horndeski Delta coefficients}
\Lambda_{i}=Q_{i},\qquad
\mathcal{J}_{ij}{}^{a}=0,\qquad
\mathcal{I}_{ij}{}^{ab}=-4F_{X^{ij}}h^{ab}.
\end{align}

Thus
\begin{align}
\label{Eq: Multi-field Horndeski Delta}
\Delta_{i}=Q_{i}-4F_{X^{ij}}S^{j},
\end{align}
with $S^{j}=h^{ab}S^{j}{}_{ab}$ as in section \ref{Sec: ADM decomposition of the action}. Since $\mathcal{W}_{iab}=0$ and $\mathcal{J}_{ij}{}^{a}=0$, Eq. \eqref{Eq: Derivative smearing coefficient} gives
\begin{align}
\label{Eq: Multi-field Horndeski D}
\mathfrak{D}_{ij}{}^{a}=0.
\end{align}

Similarly, Eq. \eqref{Eq: Momentum dependent bracket coefficient} gives
\begin{align}
\label{Eq: Multi-field Horndeski M}
\mathfrak{M}_{ij}=0,
\end{align}
because all terms in $\mathfrak{M}_{ij}$ are proportional either to $\mathcal{W}_{iab}$ or to derivatives of $\mathcal{W}_{iab}$. The only remaining primary-primary condition is the momentum-independent one. Since $\mathcal{W}_{iab}=0$ and $\mathcal{J}_{ij}{}^{a}=0$, Eq. \eqref{Eq: Momentum independent bracket coefficient} reduces to
\begin{align}
\label{Eq: Multi-field Horndeski N condition I}
\left(\frac{\partial\Delta_{i}}{\partial\mathcal{A}_{*}^{j}}-\frac{\partial\Delta_{j}}{\partial\mathcal{A}_{*}^{i}}\right)_{\rm{alg}}=0.
\end{align}

For any function $Y(\phi^{i},X^{ij})$, the algebraic derivative with respect to $\mathcal{A}_{*}^{i}$ is
\begin{align}
\label{Eq: Algebraic derivative through X}
\left(\frac{\partial Y}{\partial\mathcal{A}_{*}^{i}}\right)_{\rm{alg}}=-2\mathcal{A}_{*}^{j}Y_{X^{ij}},
\end{align}
because $X^{ij}$ depends on $\mathcal{A}_{*}^{i}$ through Eq. \eqref{Eq: Decomposition of kinetic matrix}. Applying this to Eq. \eqref{Eq: Multi-field Horndeski Delta}, and noting that $S^{i}$ is independent of $\mathcal{A}_{*}^{j}$ in the algebraic derivative, Eq. \eqref{Eq: Multi-field Horndeski N condition I} becomes
\begin{align}
\label{Eq: Multi-field Horndeski N condition II}
0=-2\mathcal{A}_{*}^{l}\left[Q_{i,X^{jl}}-Q_{j,X^{il}}-4\left(F_{X^{ik}X^{jl}}-F_{X^{jk}X^{il}}\right)S^{k}\right].
\end{align}

Here
\begin{align}
\label{Eq: Multi-field Horndeski derivative notation}
Q_{i,X^{jk}}=\frac{\partial Q_{i}}{\partial X^{jk}},\qquad
F_{X^{ij}X^{kl}}=\frac{\partial^{2}F}{\partial X^{ij}\partial X^{kl}}.
\end{align}

A sufficient coefficient-level implementation of Eq. \eqref{Eq: Multi-field Horndeski N condition II} is
\begin{align}
\label{Eq: Multi-field Horndeski closure conditions}
Q_{i,X^{jk}}=Q_{j,X^{ik}},\qquad
F_{X^{ik}X^{jl}}=F_{X^{jk}X^{il}}.
\end{align}

These are not identities satisfied by arbitrary $F$ and $Q_{i}$; they are the coefficient-level closure conditions required by the momentum-independent part of the primary-primary bracket.\footnote{Field-space symmetry conditions of this type also appear in covariant multi-Galileon \cite{Padilla:2012dx} and bi-scalar Horndeski \cite{Ohashi:2015fma} constructions, where they follow from requiring second-order field equations; here they instead arise from the momentum-independent part of the primary-primary bracket in the Hamiltonian analysis.} Hence the primary degeneracy condition is automatic in the multi-field Horndeski-type subclass, and so are the conditions $\mathfrak{D}_{ij}{}^{a}=0$ and $\mathfrak{M}_{ij}=0$, but the condition $\mathfrak{N}_{ij}=0$ imposes the nontrivial restrictions \eqref{Eq: Multi-field Horndeski closure conditions}, or a weaker solution of Eq. \eqref{Eq: Multi-field Horndeski N condition II}.
\subsection{A constructive degenerate multi-field subclass}
\label{Subsec: A constructive degenerate multi-field subclass}
The previous subsections illustrate special limits in which the general conditions simplify considerably. We now give a genuinely multi-field subclass that satisfies the full set of strong conditions while keeping nonzero scalar-metric kinetic mixing, thereby showing that the strong conditions of section \ref{Sec: Hamiltonian analysis} admit nontrivial multi-field solutions. The construction is made directly at the level of the ADM coefficient blocks in the metric-invertible branch. We do not attempt to reconstruct covariant coefficient functions $A^{(I)}$ for this subclass; the inverse map from ADM blocks to covariant coefficients is not solved here. The example should therefore be understood as an ADM-level constructive subclass.

A natural first attempt is the isotropic profile
\begin{align}
\label{Eq: Constructive isotropic profile}
\mathcal{W}_{iab}=w_{i}h_{ab},
\end{align}
with constant $w_{i}$. If $\mathcal{J}_{ij}{}^{a}=0$ and $\mathcal{I}_{ij}{}^{ab}=0$, then Eq. \eqref{Eq: R tensor definition} gives $\mathcal{R}_{i}{}^{abc}=0$, and hence Eq. \eqref{Eq: Derivative smearing coefficient} gives $\mathfrak{D}_{ij}{}^{a}=0$. Moreover, Eq. \eqref{Eq: Momentum dependent bracket coefficient} gives $\mathfrak{M}_{ij}=0$, because $\mathcal{W}_{iab}$ has no $\mathcal{A}_{*}^{i}$ dependence and its $h_{ab}$-derivative contribution has the rank-one form $w_{i}w_{j}$.

However, Eq. \eqref{Eq: Momentum independent bracket coefficient} contains the trace term
\begin{align}
\label{Eq: Constructive isotropic obstruction}
2\mathcal{W}_{[i|a|}{}^{a}\Delta_{j]}=3\left(w_{i}\Delta_{j}-w_{j}\Delta_{i}\right),
\end{align}
which does not vanish for a generic $\Delta_{i}$. For example, if $\Delta_{i}=m_{ij}\mathcal{A}_{*}^{j}$ with constant symmetric $m_{ij}$, the remaining terms in $\mathfrak{N}_{ij}$ vanish and Eq. \eqref{Eq: Constructive isotropic obstruction} imposes
\begin{align}
\label{Eq: Constructive isotropic rank one}
w_{i}m_{jk}=w_{j}m_{ik}.
\end{align}

This forces $m_{ij}$ to be proportional to $w_{i}w_{j}$, and hence rank one in field space. The isotropic profile is therefore too restrictive for a genuinely $\mathcal{N}$-field construction.

We instead keep the rank-one field-space structure, which makes $\mathfrak{M}_{ij}$ simple, but remove the trace that obstructs $\mathfrak{N}_{ij}=0$. We take
\begin{align}
\label{Eq: Constructive traceless profile}
\mathcal{W}_{iab}=w_{i}\Theta_{ab},\qquad
\Theta_{ab}=t_{jk}\hat{\mathcal{A}}^{j}{}_{a}\hat{\mathcal{A}}^{k}{}_{b}-\frac{1}{3}t_{jk}\hat{\mathcal{A}}^{jk}h_{ab},\qquad
h^{ab}\Theta_{ab}=0,
\end{align}
where $w_{i}$ is a constant vector in field space and $t_{jk}=t_{kj}$ is a constant symmetric matrix. The tensor $\Theta_{ab}$ is only a shorthand for the displayed combination of $h_{ab}$ and $\hat{\mathcal{A}}^{i}{}_{a}$; it is not a new dynamical field. It lies in the same spatial tensor span used in the general decomposition of $\mathcal{W}_{iab}$.

We choose the kinetic blocks
\begin{align}
\label{Eq: Constructive kinetic blocks}
\mathcal{B}_{i}{}^{ab}=w_{i}\mathcal{K}^{abcd}\Theta_{cd},\qquad
\mathcal{G}_{ij}=w_{i}w_{j}\mathcal{K}^{abcd}\Theta_{ab}\Theta_{cd},
\end{align}
with $\mathcal{K}^{abcd}$ invertible and satisfying the symmetries in Eq. \eqref{Eq: ADM coefficient symmetries}. We also choose
\begin{align}
\label{Eq: Constructive Delta data}
\mathcal{J}_{ij}{}^{a}=0,\qquad
\mathcal{I}_{ij}{}^{ab}=0,\qquad
\Delta_{i}=m_{ij}\mathcal{A}_{*}^{j},
\end{align}
where $m_{ij}=m_{ji}$ is a constant symmetric matrix. At the ADM-block level this choice of $\Delta_{i}$ can be implemented by choosing
\begin{align}
\label{Eq: Constructive C choice}
\mathcal{C}_{i}=\mathcal{W}_{iab}\mathcal{M}^{ab}-m_{ij}\mathcal{A}_{*}^{j}.
\end{align}

The field-space mixing is carried by $t_{ij}$ and $m_{ij}$, and $m_{ij}$ may be chosen to be full rank. The same proof remains valid if $w_{i}$, $t_{ij}$ and $m_{ij}$ are promoted to functions of $\phi^{k}$, since the strong conditions involve algebraic derivatives with respect to $\mathcal{A}_{*}^{i}$ and $h_{ab}$.

The primary degeneracy condition is satisfied by construction. Indeed, using Eq. \eqref{Eq: Constructive kinetic blocks},
\begin{align}
\label{Eq: Constructive primary degeneracy}
\mathcal{S}_{ij}=\mathcal{G}_{ij}-\mathcal{B}_{i}{}^{ab}\mathcal{W}_{jab}=w_{i}w_{j}\mathcal{K}^{abcd}\Theta_{ab}\Theta_{cd}-w_{i}w_{j}\mathcal{K}^{abcd}\Theta_{cd}\Theta_{ab}=0.
\end{align}

The derivative-smearing condition also vanishes. Since $\mathcal{I}_{ij}{}^{ab}=0$, Eq. \eqref{Eq: R tensor definition} gives
\begin{align}
\label{Eq: Constructive R vanishes}
\mathcal{R}_{i}{}^{abc}=0.
\end{align}

Together with $\mathcal{J}_{ij}{}^{a}=0$, Eq. \eqref{Eq: Derivative smearing coefficient} gives
\begin{align}
\label{Eq: Constructive D vanishes}
\mathfrak{D}_{ij}{}^{a}=0.
\end{align}

The momentum-dependent condition follows from the rank-one form of $\mathcal{W}_{iab}$. Since $\Theta_{ab}$ has no $\mathcal{A}_{*}^{i}$ dependence, the first term in Eq. \eqref{Eq: Momentum dependent bracket coefficient} vanishes. The second term is
\begin{align}
\label{Eq: Constructive M vanishes}
&\left(\frac{\partial\mathcal{W}_{iab}}{\partial h_{cd}}\right)_{\rm{alg}}\mathcal{W}_{jcd}-\mathcal{W}_{icd}\left(\frac{\partial\mathcal{W}_{jab}}{\partial h_{cd}}\right)_{\rm{alg}}\nonumber\\&=w_{i}w_{j}\left[\left(\frac{\partial\Theta_{ab}}{\partial h_{cd}}\right)_{\rm{alg}}\Theta_{cd}-\Theta_{cd}\left(\frac{\partial\Theta_{ab}}{\partial h_{cd}}\right)_{\rm{alg}}\right]=0.
\end{align}

Therefore
\begin{align}
\label{Eq: Constructive M zero}
\mathfrak{M}_{ij}=0.
\end{align}

Finally, the momentum-independent condition vanishes because of the symmetry of $m_{ij}$ and the tracelessness of $\Theta_{ab}$. The last term in Eq. \eqref{Eq: Momentum independent bracket coefficient} vanishes because $\mathcal{R}_{i}{}^{abc}=0$, and the term $-2D_{a}\mathcal{J}_{[ij]}{}^{a}$ vanishes because $\mathcal{J}_{ij}{}^{a}=0$. Moreover,
\begin{align}
\label{Eq: Constructive N derivative pieces}
\left(\frac{\partial\Delta_{i}}{\partial\mathcal{A}_{*}^{j}}-\frac{\partial\Delta_{j}}{\partial\mathcal{A}_{*}^{i}}\right)_{\rm{alg}}=m_{ij}-m_{ji}=0,\qquad
\left(\frac{\partial\Delta_{i}}{\partial h_{ab}}\right)_{\rm{alg}}=0.
\end{align}

The remaining trace term is
\begin{align}
\label{Eq: Constructive N trace}
2\mathcal{W}_{[i|a|}{}^{a}\Delta_{j]}=w_{i}\left(h^{ab}\Theta_{ab}\right)\Delta_{j}-w_{j}\left(h^{ab}\Theta_{ab}\right)\Delta_{i}=0.
\end{align}

Thus
\begin{align}
\label{Eq: Constructive N zero}
\mathfrak{N}_{ij}=0.
\end{align}

Equations \eqref{Eq: Constructive primary degeneracy}, \eqref{Eq: Constructive D vanishes}, \eqref{Eq: Constructive M zero} and \eqref{Eq: Constructive N zero} show that the subclass defined by Eqs. \eqref{Eq: Constructive traceless profile}--\eqref{Eq: Constructive Delta data} satisfies the complete primary degeneracy condition and the strong primary-primary conditions \eqref{Eq: Strong primary primary conditions}. Since $\mathcal{B}_{i}{}^{ab}$ is nonzero, this subclass is not of the multi-field Horndeski type discussed in subsection \ref{Subsec: A multi-field quadratic Horndeski-type subclass}. For $\mathcal{N}\ge2$, suitable choices of $t_{ij}$ and $m_{ij}$ make the subclass genuinely multi-field rather than a collection of independent single-field sectors. The corresponding primary constraints are
\begin{align}
\label{Eq: Constructive primary constraints}
\Psi_{i}=p_{*i}-2w_{i}\Theta_{ab}\pi^{ab}+\sqrt{h}m_{ij}\mathcal{A}_{*}^{j}\approx0.
\end{align}

The conditions verified above are those of the primary degeneracy and primary-primary consistency analysis. As in the general analysis, if in addition the rank condition \eqref{Eq: Rank condition} is satisfied, the scalar primary-secondary chain closes in the desired way and the constraints $\Psi_{i}$ and $\Omega_{i}$ form $\mathcal{N}$ second-class pairs in the scalar degeneracy sector. The degree-of-freedom count of subsection \ref{Subsec: Counting degrees of freedom} then gives $2+\mathcal{N}$ propagating degrees of freedom, under the full set of assumptions stated there, in particular that this scalar second-class sector leaves the diffeomorphism constraints first class. Neither the rank condition \eqref{Eq: Rank condition} nor the first-class character of the diffeomorphism sector is verified by the ADM-level construction given here. What this subclass establishes is that the Hamiltonian degeneracy and primary-primary consistency conditions of section \ref{Sec: Hamiltonian analysis} admit a nontrivial solution with nonzero scalar-metric kinetic mixing.
\section{Discussion and conclusion}
\label{Sec: Discussion and conclusion}
In this work we developed a Hamiltonian framework for removing the Ostrogradsky modes of multi-field higher-order scalar-tensor theories, and derived explicit degeneracy and consistency conditions on the coefficient functions. We considered $\mathcal{N}$ scalar fields with coefficient functions of $\phi^{i}$ and the kinetic matrix $X^{ij}$, and with quadratic dependence on the second derivatives of the fields. We introduced auxiliary co-vectors $\mathcal{A}^{i}{}_{\mu}$ and performed an ADM decomposition, which isolates the velocity sector built from the scalar velocities $\mathcal{V}_{*}^{i}$ and the metric velocity $K_{ab}$ and lets us formulate the degeneracy problem directly in terms of the ADM coefficient blocks. The degree-of-freedom count also relies on the assumptions stated in subsection \ref{Subsec: Counting degrees of freedom}.

In the metric-invertible branch, where $\mathcal{K}^{abcd}$ is invertible, the scalar-metric kinetic mixing is carried by the auxiliary tensor $\mathcal{W}_{iab}$, and complete primary degeneracy is the vanishing of the Schur complement, $\mathcal{S}_{ij}=0$, Eq. \eqref{Eq: Degeneracy condition}. When this condition holds, the Legendre map is degenerate and one obtains $\mathcal{N}$ primary constraints $\Psi_{i}$, Eq. \eqref{Eq: Primary degeneracy constraints}, the multi-field generalization of the primary degeneracy constraint familiar from the single-field case.

The difference from the single-field case appears at the next step. For $\mathcal{N}\ge2$, preservation of the primary constraints produces new conditions on the coefficient functions. After eliminating the auxiliary second-class pair $(\chi^{i}{}_{a},\hat{p}_{i}{}^{a})$, the smeared primary-primary bracket separates into a derivative-smearing coefficient $\mathfrak{D}_{ij}{}^{a}$ and the non-derivative coefficients $\mathfrak{M}_{ij}$ and $\mathfrak{N}_{ij}$. The latter two are antisymmetric in the field-space indices and therefore have no single-field analogue. For the primary constraints to generate secondary constraints, rather than to fix their own multipliers, all three coefficients must vanish on the constraint surface, Eq. \eqref{Eq: Weak primary primary conditions}. That coupled higher-derivative systems can require conditions beyond primary degeneracy was already known \cite{Motohashi:2016ftl,Klein:2016aiq}, and the field-theory analysis of \cite{Crisostomi:2017aim} organized the secondary conditions into symmetric and antisymmetric parts in the field-space indices, under assumptions that exclude gauge symmetries and a dynamical spin-2 field. The present work realizes this constraint-preservation mechanism in a generally covariant multi-scalar-tensor theory with a dynamical metric and arbitrary $\mathcal{N}$, where the scalar and metric velocities mix in the ADM Hessian and the gravitational constraint sector is present, and obtains the consistency coefficients explicitly in terms of the ADM quantities generated by the covariant action.

Once the conditions \eqref{Eq: Weak primary primary conditions} hold, preservation of $\Psi_{i}$ gives $\mathcal{N}$ secondary constraints $\Omega_{i}$. If the primary-secondary bracket has maximal rank, Eq. \eqref{Eq: Rank condition}, the pairs $(\Psi_{i},\Omega_{i})$ are second class in the scalar degeneracy sector and remove the unwanted scalar phase-space directions. Under the assumptions of subsection \ref{Subsec: Counting degrees of freedom}, and in particular under the assumption that this scalar second-class sector leaves the diffeomorphism constraints first class, the number of propagating degrees of freedom is $N_{\rm{dof}}=2+\mathcal{N}$, Eq. \eqref{Eq: Final number of degrees of freedom}, the two tensor modes of gravity together with one scalar mode per field, and no additional Ostrogradsky mode.

We examined the construction in several special cases. In the single-field limit, the antisymmetric coefficients $\mathfrak{M}_{ij}$ and $\mathfrak{N}_{ij}$ vanish identically, while the derivative-smearing coefficient $\mathfrak{D}_{ij}{}^{a}$ vanishes only after the single-field degeneracy conditions are imposed. In the single-field case the primary-primary bracket therefore imposes no additional condition beyond primary degeneracy, and the reduction of the degeneracy condition provides a check against the degeneracy structure known for single-field quadratic higher-order scalar-tensor theories \cite{Langlois:2015skt}. In the multi-field quadratic Horndeski-type subclass of subsection \ref{Subsec: A multi-field quadratic Horndeski-type subclass}, primary degeneracy is automatic in the metric-invertible branch, the conditions $\mathfrak{D}_{ij}{}^{a}=0$ and $\mathfrak{M}_{ij}=0$ hold identically, whereas $\mathfrak{N}_{ij}=0$ imposes a nontrivial closure condition on the coefficient functions. The multi-field problem is therefore more restrictive than a term-by-term extension of the single-field case.

We also gave a constructive multi-field subclass with nonvanishing scalar-metric kinetic mixing, $\mathcal{W}_{iab}\neq0$, defined directly at the level of the ADM coefficient blocks. It satisfies complete primary degeneracy and the strong form of the primary-primary conditions, Eq. \eqref{Eq: Strong primary primary conditions}, so these conditions admit a nontrivial solution beyond the single-field and Horndeski-type limits. We do not reconstruct covariant coefficient functions $A^{(I)}$ for this subclass, and we verify neither the rank condition \eqref{Eq: Rank condition} nor the assumption that the scalar second-class sector leaves the diffeomorphism constraints first class. It therefore does not by itself establish a $2+\mathcal{N}$ count, and should be read as an ADM-level demonstration that the Hamiltonian degeneracy and primary-primary consistency conditions admit nontrivial solutions, not as a covariant model-building prescription.

The present analysis is restricted to the metric-invertible branch. Branches in which the metric block is itself degenerate may give further classes of theories and call for a separate constraint analysis. A covariant realization of the ADM-level subclass remains to be constructed. Its perturbative dynamics and cosmological applications can then be investigated. In explicit covariant representatives, two ingredients of the counting remain to be verified: the primary-secondary rank condition \eqref{Eq: Rank condition}, since accidental degeneracies may produce additional constraints or further gauge symmetries, and the assumption that the scalar second-class sector leaves the diffeomorphism constraints first class. Under the stated rank and constraint-algebra assumptions, the explicit degeneracy and primary-primary consistency conditions derived here remove the unwanted higher-order scalar modes in the metric-invertible branch. The analysis does not give a classification of healthy multi-field theories. These Hamiltonian conditions can be applied to the construction and analysis of genuinely multi-field higher-order scalar-tensor theories beyond direct single-field generalizations.
\appendix
\section{Explicit coefficients in the ADM action}
\label{AppSec: Explicit coefficients in the ADM action}
In this appendix we collect the explicit coefficient functions appearing in the ADM action \eqref{Eq: ADM action}. All quantities are written in terms of the ADM variables introduced in section \ref{Sec: ADM decomposition of the action}, and their symmetry properties are those stated in Eq. \eqref{Eq: ADM coefficient symmetries}.

The term linear in the acceleration is controlled by $\mathcal{E}^{a}=h^{ab}\mathcal{E}_{b}$, with
\begin{align}
\label{Eq: Acceleration coefficient in ADM action}
\mathcal{E}_{a}=2F_{\phi^{i}}\hat{\mathcal{A}}^{i}{}_{a}+4F_{X^{ij}}\left(\hat{\mathcal{A}}^{ib}S^{j}{}_{ab}-\mathcal{A}_{*}^{i}D_{a}\mathcal{A}_{*}^{j}\right).
\end{align}

The part independent of $K_{ab}$, $\mathcal{V}_{*}^{i}$, and explicit acceleration is
\begin{align}
\label{Eq: Potential coefficient in ADM action}
\mathcal{U}&=P+Q_{i}S^{i}+B_{ijk}\hat{\mathcal{A}}^{ia}\hat{\mathcal{A}}^{jb}S^{k}{}_{ab}-2B_{ijk}\hat{\mathcal{A}}^{ia}\mathcal{A}_{*}^{j}D_{a}\mathcal{A}_{*}^{k}+A^{(1)}_{ij}S^{i}{}_{ab}S^{jab}-2A^{(1)}_{ij}D_{a}\mathcal{A}_{*}^{j}D^{a}\mathcal{A}_{*}^{i}\nonumber\\&+A^{(2)}_{ij}S^{i}S^{j}+A^{(3)}_{ijlk}\hat{\mathcal{A}}^{ia}\hat{\mathcal{A}}^{jb}S^{l}{}_{ab}S^{k}-2A^{(3)}_{ijlk}\hat{\mathcal{A}}^{ia}\mathcal{A}_{*}^{j}S^{k}D_{a}\mathcal{A}_{*}^{l}+A^{(4)}_{ijkl}\hat{\mathcal{A}}^{ia}\hat{\mathcal{A}}^{jb}S^{k}{}_{ac}S^{l}{}_{b}{}^{c}\nonumber\\&+A^{(4)}_{ijkl}\mathcal{A}_{*}^{i}\mathcal{A}_{*}^{j}D_{a}\mathcal{A}_{*}^{l}D^{a}\mathcal{A}_{*}^{k}-A^{(4)}_{ijkl}\hat{\mathcal{A}}^{ia}\hat{\mathcal{A}}^{jb}D_{a}\mathcal{A}_{*}^{k}D_{b}\mathcal{A}_{*}^{l}-2A^{(4)}_{ijlk}\hat{\mathcal{A}}^{ia}\mathcal{A}_{*}^{j}S^{l}{}_{ab}D^{b}\mathcal{A}_{*}^{k}\nonumber\\&+A^{(5)}_{ijklmn}\hat{\mathcal{A}}^{ia}\hat{\mathcal{A}}^{jb}\hat{\mathcal{A}}^{kc}\hat{\mathcal{A}}^{ld}S^{m}{}_{ab}S^{n}{}_{cd}-4A^{(5)}_{iljkmn}\hat{\mathcal{A}}^{ia}\hat{\mathcal{A}}^{jb}\hat{\mathcal{A}}^{kc}\mathcal{A}_{*}^{l}S^{n}{}_{bc}D_{a}\mathcal{A}_{*}^{m}\nonumber\\&+4A^{(5)}_{ikjlmn}\hat{\mathcal{A}}^{ia}\hat{\mathcal{A}}^{jb}\mathcal{A}_{*}^{k}\mathcal{A}_{*}^{l}D_{a}\mathcal{A}_{*}^{m}D_{b}\mathcal{A}_{*}^{n}.
\end{align}

The term linear in $\mathcal{V}_{*}^{i}$ is determined by
\begin{align}
\mathcal{C}_{i}&=-Q_{i}+B_{jki}\mathcal{A}_{*}^{j}\mathcal{A}_{*}^{k}-2A^{(2)}_{ij}S^{j}-A^{(3)}_{jkli}\hat{\mathcal{A}}^{ja}\hat{\mathcal{A}}^{kb}S^{l}{}_{ab}+A^{(3)}_{jkil}\mathcal{A}_{*}^{j}\mathcal{A}_{*}^{k}S^{l}+2A^{(3)}_{jkli}\hat{\mathcal{A}}^{ja}\mathcal{A}_{*}^{k}D_{a}\mathcal{A}_{*}^{l}\nonumber\\&+2A^{(4)}_{kjil}\hat{\mathcal{A}}^{ja}\mathcal{A}_{*}^{k}D_{a}\mathcal{A}_{*}^{l}+2A^{(5)}_{lmjkin}\hat{\mathcal{A}}^{ja}\hat{\mathcal{A}}^{kb}\mathcal{A}_{*}^{l}\mathcal{A}_{*}^{m}S^{n}{}_{ab}-4A^{(5)}_{kljmin}\hat{\mathcal{A}}^{ja}\mathcal{A}_{*}^{k}\mathcal{A}_{*}^{l}\mathcal{A}_{*}^{m}D_{a}\mathcal{A}_{*}^{n}.\nonumber\\&
\label{Eq: Auxiliary velocity coefficient in ADM action}
\end{align}

The term linear in $K_{ab}$ is determined by
\begin{align}
\label{Eq: Extrinsic curvature coefficient in ADM action}
\mathcal{M}^{ab}&=-2F_{\phi^{i}}\mathcal{A}_{*}^{i}h^{ab}-4F_{X^{ij}}\hat{\mathcal{A}}^{ic}h^{ab}D_{c}\mathcal{A}_{*}^{j}-Q_{i}\mathcal{A}_{*}^{i}h^{ab}-B_{ijk}\hat{\mathcal{A}}^{ia}\hat{\mathcal{A}}^{jb}\mathcal{A}_{*}^{k}+2B_{ikj}\hat{\mathcal{A}}^{i(a}\hat{\mathcal{A}}^{|j|b)}\mathcal{A}_{*}^{k}\nonumber\\&-2A^{(1)}_{ij}\mathcal{A}_{*}^{i}S^{jab}+4A^{(1)}_{ij}\hat{\mathcal{A}}^{i(a}D^{b)}\mathcal{A}_{*}^{j}-2A^{(2)}_{ij}\mathcal{A}_{*}^{i}h^{ab}S^{j}-A^{(3)}_{ijlk}\hat{\mathcal{A}}^{ic}\hat{\mathcal{A}}^{jd}\mathcal{A}_{*}^{k}h^{ab}S^{l}{}_{cd}\nonumber\\&+2A^{(3)}_{ijlk}\hat{\mathcal{A}}^{ic}\mathcal{A}_{*}^{j}\mathcal{A}_{*}^{k}h^{ab}D_{c}\mathcal{A}_{*}^{l}-A^{(3)}_{ijkl}\hat{\mathcal{A}}^{ia}\hat{\mathcal{A}}^{jb}\mathcal{A}_{*}^{k}S^{l}+A^{(3)}_{ijkl}\hat{\mathcal{A}}^{i(a}\hat{\mathcal{A}}^{|k|b)}\mathcal{A}_{*}^{j}S^{l}\nonumber\\&+A^{(3)}_{ijkl}\hat{\mathcal{A}}^{k(a}\hat{\mathcal{A}}^{|i|b)}\mathcal{A}_{*}^{j}S^{l}+A^{(4)}_{ijkl}\hat{\mathcal{A}}^{i(a}\hat{\mathcal{A}}^{|k|b)}\hat{\mathcal{A}}^{jc}D_{c}\mathcal{A}_{*}^{l}+A^{(4)}_{ijkl}\hat{\mathcal{A}}^{k(a}\hat{\mathcal{A}}^{|i|b)}\hat{\mathcal{A}}^{jc}D_{c}\mathcal{A}_{*}^{l}\nonumber\\&-2A^{(4)}_{ijkl}\hat{\mathcal{A}}^{i(a}\hat{\mathcal{A}}^{|jc}\mathcal{A}_{*}^{k}S^{l|b)}{}_{c}+2A^{(4)}_{ijkl}\hat{\mathcal{A}}^{k(a}\hat{\mathcal{A}}^{|jc}\mathcal{A}_{*}^{i}S^{l|b)}{}_{c}+2A^{(4)}_{ijkl}\hat{\mathcal{A}}^{i(a}\mathcal{A}_{*}^{|j}\mathcal{A}_{*}^{k|}D^{b)}\mathcal{A}_{*}^{l}\nonumber\\&-2A^{(4)}_{ijkl}\hat{\mathcal{A}}^{k(a}\mathcal{A}_{*}^{|i}\mathcal{A}_{*}^{j|}D^{b)}\mathcal{A}_{*}^{l}-2A^{(5)}_{ijklmn}\hat{\mathcal{A}}^{ia}\hat{\mathcal{A}}^{jb}\hat{\mathcal{A}}^{kc}\hat{\mathcal{A}}^{ld}\mathcal{A}_{*}^{m}S^{n}{}_{cd}\nonumber\\&+2A^{(5)}_{ijklmn}\hat{\mathcal{A}}^{i(a}\hat{\mathcal{A}}^{|m|b)}\hat{\mathcal{A}}^{kc}\hat{\mathcal{A}}^{ld}\mathcal{A}_{*}^{j}S^{n}{}_{cd}+2A^{(5)}_{ijklmn}\hat{\mathcal{A}}^{m(a}\hat{\mathcal{A}}^{|i|b)}\hat{\mathcal{A}}^{kc}\hat{\mathcal{A}}^{ld}\mathcal{A}_{*}^{j}S^{n}{}_{cd}\nonumber\\&+4A^{(5)}_{ijklmn}\hat{\mathcal{A}}^{ia}\hat{\mathcal{A}}^{jb}\hat{\mathcal{A}}^{kc}\mathcal{A}_{*}^{l}\mathcal{A}_{*}^{m}D_{c}\mathcal{A}_{*}^{n}-4A^{(5)}_{ijklmn}\hat{\mathcal{A}}^{i(a}\hat{\mathcal{A}}^{|m|b)}\hat{\mathcal{A}}^{kc}\mathcal{A}_{*}^{j}\mathcal{A}_{*}^{l}D_{c}\mathcal{A}_{*}^{n}\nonumber\\&-4A^{(5)}_{ijklmn}\hat{\mathcal{A}}^{m(a}\hat{\mathcal{A}}^{|i|b)}\hat{\mathcal{A}}^{kc}\mathcal{A}_{*}^{j}\mathcal{A}_{*}^{l}D_{c}\mathcal{A}_{*}^{n}.
\end{align}

The coefficients quadratic in the velocities $\mathcal{V}_{*}^{i}$ and $K_{ab}$ are as follows. The coefficient of $\mathcal{V}_{*}^{i}\mathcal{V}_{*}^{j}$ is
\begin{align}
\label{Eq: Auxiliary velocity Hessian coefficient in ADM action}
\mathcal{G}_{ij}=A^{(1)}_{ij}+A^{(2)}_{ij}-A^{(3)}_{kl(ij)}\mathcal{A}_{*}^{k}\mathcal{A}_{*}^{l}-A^{(4)}_{klij}\mathcal{A}_{*}^{k}\mathcal{A}_{*}^{l}+A^{(5)}_{klmnij}\mathcal{A}_{*}^{k}\mathcal{A}_{*}^{l}\mathcal{A}_{*}^{m}\mathcal{A}_{*}^{n}.
\end{align}

The coefficient of $\mathcal{V}_{*}^{i}K_{ab}$ is
\begin{align}
\label{Eq: Mixed kinetic coefficient in ADM action}
\mathcal{B}_{i}{}^{ab}&=2F_{X^{ij}}\mathcal{A}_{*}^{j}h^{ab}+A^{(2)}_{ij}\mathcal{A}_{*}^{j}h^{ab}-\frac{1}{2}A^{(3)}_{jkil}\mathcal{A}_{*}^{j}\mathcal{A}_{*}^{k}\mathcal{A}_{*}^{l}h^{ab}+\frac{1}{2}A^{(3)}_{jkli}\hat{\mathcal{A}}^{ja}\hat{\mathcal{A}}^{kb}\mathcal{A}_{*}^{l}\nonumber\\&-\frac{1}{2}A^{(3)}_{jkli}\hat{\mathcal{A}}^{j(a}\hat{\mathcal{A}}^{|l|b)}\mathcal{A}_{*}^{k}-\frac{1}{2}A^{(3)}_{jkli}\hat{\mathcal{A}}^{l(a}\hat{\mathcal{A}}^{|j|b)}\mathcal{A}_{*}^{k}-\frac{1}{2}A^{(4)}_{jkil}\hat{\mathcal{A}}^{k(a}\hat{\mathcal{A}}^{|l|b)}\mathcal{A}_{*}^{j}\nonumber\\&-\frac{1}{2}A^{(4)}_{jkil}\hat{\mathcal{A}}^{l(a}\hat{\mathcal{A}}^{|k|b)}\mathcal{A}_{*}^{j}-A^{(5)}_{jklmin}\hat{\mathcal{A}}^{la}\hat{\mathcal{A}}^{mb}\mathcal{A}_{*}^{j}\mathcal{A}_{*}^{k}\mathcal{A}_{*}^{n}+A^{(5)}_{jklmin}\hat{\mathcal{A}}^{l(a}\hat{\mathcal{A}}^{|n|b)}\mathcal{A}_{*}^{j}\mathcal{A}_{*}^{k}\mathcal{A}_{*}^{m}\nonumber\\&+A^{(5)}_{jklmin}\hat{\mathcal{A}}^{n(a}\hat{\mathcal{A}}^{|l|b)}\mathcal{A}_{*}^{j}\mathcal{A}_{*}^{k}\mathcal{A}_{*}^{m}.
\end{align}

For compactness, the coefficient of $K_{ab}K_{cd}$ is displayed with two indices lowered, namely
\begin{align}
\label{Eq: Mixed index definition of K coefficient}
\mathcal{K}^{ab}{}_{cd}=\mathcal{K}^{abef}h_{ec}h_{fd}.
\end{align}

Equivalently, the term $\mathcal{K}^{abcd}K_{ab}K_{cd}$ in Eq. \eqref{Eq: ADM action} can be written as $\mathcal{K}^{ab}{}_{cd}K_{ab}K^{cd}$. The explicit mixed-index expression is
\begin{align}
\label{Eq: Extrinsic curvature Hessian coefficient in ADM action}
\mathcal{K}^{ab}{}_{cd}&=-Fh^{ab}h_{cd}+F\delta^{a}{}_{(c}\delta^{b}{}_{d)}+2F_{X^{ij}}\hat{\mathcal{A}}^{ia}\hat{\mathcal{A}}^{jb}h_{cd}+2F_{X^{ij}}\hat{\mathcal{A}}^{i}{}_{c}\hat{\mathcal{A}}^{j}{}_{d}h^{ab}+A^{(1)}_{ij}\mathcal{A}_{*}^{i}\mathcal{A}_{*}^{j}\delta^{a}{}_{(c}\delta^{b}{}_{d)}\nonumber\\&-2A^{(1)}_{ij}\hat{\mathcal{A}}^{i(a}\hat{\mathcal{A}}^{|j|}{}_{(c}\delta_{d)}{}^{b)}+A^{(2)}_{ij}\mathcal{A}_{*}^{i}\mathcal{A}_{*}^{j}h^{ab}h_{cd}+\frac{1}{2}A^{(3)}_{ijkl}\hat{\mathcal{A}}^{ia}\hat{\mathcal{A}}^{jb}\mathcal{A}_{*}^{k}\mathcal{A}_{*}^{l}h_{cd}\nonumber\\&-\frac{1}{2}A^{(3)}_{ijkl}\hat{\mathcal{A}}^{i(a}\hat{\mathcal{A}}^{|k|b)}\mathcal{A}_{*}^{j}\mathcal{A}_{*}^{l}h_{cd}+\frac{1}{2}A^{(3)}_{ijkl}\hat{\mathcal{A}}^{i}{}_{c}\hat{\mathcal{A}}^{j}{}_{d}\mathcal{A}_{*}^{k}\mathcal{A}_{*}^{l}h^{ab}-\frac{1}{2}A^{(3)}_{ijkl}\hat{\mathcal{A}}^{i}{}_{(c}\hat{\mathcal{A}}^{k}{}_{d)}\mathcal{A}_{*}^{j}\mathcal{A}_{*}^{l}h^{ab}\nonumber\\&-\frac{1}{2}A^{(3)}_{ijkl}\hat{\mathcal{A}}^{k(a}\hat{\mathcal{A}}^{|i|b)}\mathcal{A}_{*}^{j}\mathcal{A}_{*}^{l}h_{cd}-\frac{1}{2}A^{(3)}_{ijkl}\hat{\mathcal{A}}^{k}{}_{(c}\hat{\mathcal{A}}^{i}{}_{d)}\mathcal{A}_{*}^{j}\mathcal{A}_{*}^{l}h^{ab}-\frac{1}{4}A^{(4)}_{ijkl}\hat{\mathcal{A}}^{i(a}\hat{\mathcal{A}}^{|k|b)}\hat{\mathcal{A}}^{j}{}_{(c}\hat{\mathcal{A}}^{l}{}_{d)}\nonumber\\&-\frac{1}{4}A^{(4)}_{ijkl}\hat{\mathcal{A}}^{i(a}\hat{\mathcal{A}}^{|k|b)}\hat{\mathcal{A}}^{l}{}_{(c}\hat{\mathcal{A}}^{j}{}_{d)}-\frac{1}{4}A^{(4)}_{ijkl}\hat{\mathcal{A}}^{k(a}\hat{\mathcal{A}}^{|i|b)}\hat{\mathcal{A}}^{j}{}_{(c}\hat{\mathcal{A}}^{l}{}_{d)}-\frac{1}{4}A^{(4)}_{ijkl}\hat{\mathcal{A}}^{k(a}\hat{\mathcal{A}}^{|i|b)}\hat{\mathcal{A}}^{l}{}_{(c}\hat{\mathcal{A}}^{j}{}_{d)}\nonumber\\&+A^{(4)}_{ijkl}\hat{\mathcal{A}}^{i(a}\hat{\mathcal{A}}^{|j}{}_{(c}\mathcal{A}_{*}^{k}\mathcal{A}_{*}^{l|}\delta_{d)}{}^{b)}-A^{(4)}_{ijkl}\hat{\mathcal{A}}^{i(a}\hat{\mathcal{A}}^{|l}{}_{(c}\mathcal{A}_{*}^{j}\mathcal{A}_{*}^{k|}\delta_{d)}{}^{b)}-A^{(4)}_{ijkl}\hat{\mathcal{A}}^{k(a}\hat{\mathcal{A}}^{|j}{}_{(c}\mathcal{A}_{*}^{i}\mathcal{A}_{*}^{l|}\delta_{d)}{}^{b)}\nonumber\\&+A^{(4)}_{ijkl}\hat{\mathcal{A}}^{k(a}\hat{\mathcal{A}}^{|l}{}_{(c}\mathcal{A}_{*}^{i}\mathcal{A}_{*}^{j|}\delta_{d)}{}^{b)}+A^{(5)}_{ijklmn}\hat{\mathcal{A}}^{ia}\hat{\mathcal{A}}^{jb}\hat{\mathcal{A}}^{k}{}_{c}\hat{\mathcal{A}}^{l}{}_{d}\mathcal{A}_{*}^{m}\mathcal{A}_{*}^{n}\nonumber\\&-A^{(5)}_{ijklmn}\hat{\mathcal{A}}^{ia}\hat{\mathcal{A}}^{jb}\hat{\mathcal{A}}^{k}{}_{(c}\hat{\mathcal{A}}^{n}{}_{d)}\mathcal{A}_{*}^{l}\mathcal{A}_{*}^{m}-A^{(5)}_{ijklmn}\hat{\mathcal{A}}^{ia}\hat{\mathcal{A}}^{jb}\hat{\mathcal{A}}^{n}{}_{(c}\hat{\mathcal{A}}^{k}{}_{d)}\mathcal{A}_{*}^{l}\mathcal{A}_{*}^{m}\nonumber\\&-A^{(5)}_{ijklmn}\hat{\mathcal{A}}^{i(a}\hat{\mathcal{A}}^{|m|b)}\hat{\mathcal{A}}^{k}{}_{c}\hat{\mathcal{A}}^{l}{}_{d}\mathcal{A}_{*}^{j}\mathcal{A}_{*}^{n}+A^{(5)}_{ijklmn}\hat{\mathcal{A}}^{i(a}\hat{\mathcal{A}}^{|m|b)}\hat{\mathcal{A}}^{k}{}_{(c}\hat{\mathcal{A}}^{n}{}_{d)}\mathcal{A}_{*}^{j}\mathcal{A}_{*}^{l}\nonumber\\&+A^{(5)}_{ijklmn}\hat{\mathcal{A}}^{i(a}\hat{\mathcal{A}}^{|m|b)}\hat{\mathcal{A}}^{n}{}_{(c}\hat{\mathcal{A}}^{k}{}_{d)}\mathcal{A}_{*}^{j}\mathcal{A}_{*}^{l}-A^{(5)}_{ijklmn}\hat{\mathcal{A}}^{m(a}\hat{\mathcal{A}}^{|i|b)}\hat{\mathcal{A}}^{k}{}_{c}\hat{\mathcal{A}}^{l}{}_{d}\mathcal{A}_{*}^{j}\mathcal{A}_{*}^{n}\nonumber\\&+A^{(5)}_{ijklmn}\hat{\mathcal{A}}^{m(a}\hat{\mathcal{A}}^{|i|b)}\hat{\mathcal{A}}^{k}{}_{(c}\hat{\mathcal{A}}^{n}{}_{d)}\mathcal{A}_{*}^{j}\mathcal{A}_{*}^{l}+A^{(5)}_{ijklmn}\hat{\mathcal{A}}^{m(a}\hat{\mathcal{A}}^{|i|b)}\hat{\mathcal{A}}^{n}{}_{(c}\hat{\mathcal{A}}^{k}{}_{d)}\mathcal{A}_{*}^{j}\mathcal{A}_{*}^{l}.
\end{align}
\section{Coefficient system for the metric-invertible branch}
\label{AppSec: Coefficient system for the metric-invertible branch}
In this appendix we collect the explicit coefficient functions used in section \ref{Sec: Complete degeneracy in the metric-invertible branch}. The blocks $\mathsf{A}$, $\mathsf{B}^{ij}$, $\mathsf{C}_{ij}$ and $\mathsf{D}_{ij}{}^{kl}$ in the decomposition \eqref{Eq: LHS decomposition} are
\begin{align}
\mathsf{A}=-2F+2F_{X^{ij}}\hat{\mathcal{A}}^{ij}+A^{(1)}_{ij}\mathcal{A}_{*}^{i}\mathcal{A}_{*}^{j}+3A^{(2)}_{ij}\mathcal{A}_{*}^{i}\mathcal{A}_{*}^{j}-A^{(3)}_{iklj}\hat{\mathcal{A}}^{kl}\mathcal{A}_{*}^{i}\mathcal{A}_{*}^{j}+\frac{1}{2}A^{(3)}_{klij}\hat{\mathcal{A}}^{kl}\mathcal{A}_{*}^{i}\mathcal{A}_{*}^{j},\nonumber\\&
\label{Eq: Appendix LHS coefficients II}
\end{align}
\begin{align}
\label{Eq: Appendix LHS coefficients III}
\mathsf{B}^{ij}&=-F\hat{\mathcal{A}}^{ij}+2F_{X^{kl}}\hat{\mathcal{A}}^{ik}\hat{\mathcal{A}}^{jl}+A^{(2)}_{kl}\hat{\mathcal{A}}^{ij}\mathcal{A}_{*}^{k}\mathcal{A}_{*}^{l}+\frac{1}{2}A^{(3)}_{klmn}\hat{\mathcal{A}}^{ik}\hat{\mathcal{A}}^{jl}\mathcal{A}_{*}^{m}\mathcal{A}_{*}^{n}\nonumber\\&-A^{(3)}_{klmn}\hat{\mathcal{A}}^{(i|k|}\hat{\mathcal{A}}^{j)m}\mathcal{A}_{*}^{l}\mathcal{A}_{*}^{n},
\end{align}
\begin{align}
\label{Eq: Appendix LHS coefficients IV}
\mathsf{C}_{ij}&=6F_{X^{ij}}-2A^{(1)}_{ij}+\frac{3}{2}A^{(3)}_{ijkl}\mathcal{A}_{*}^{k}\mathcal{A}_{*}^{l}-3A^{(3)}_{(i|k|j)l}\mathcal{A}_{*}^{k}\mathcal{A}_{*}^{l}-A^{(4)}_{(i|k|j)l}\hat{\mathcal{A}}^{kl}+A^{(4)}_{ijkl}\mathcal{A}_{*}^{k}\mathcal{A}_{*}^{l}\nonumber\\&-2A^{(4)}_{(i|kl|j)}\mathcal{A}_{*}^{k}\mathcal{A}_{*}^{l}+A^{(4)}_{klij}\mathcal{A}_{*}^{k}\mathcal{A}_{*}^{l}-2A^{(5)}_{ijkmln}\hat{\mathcal{A}}^{mn}\mathcal{A}_{*}^{k}\mathcal{A}_{*}^{l}+A^{(5)}_{ijmnkl}\hat{\mathcal{A}}^{mn}\mathcal{A}_{*}^{k}\mathcal{A}_{*}^{l}\nonumber\\&-2A^{(5)}_{(i|klm|j)n}\hat{\mathcal{A}}^{lm}\mathcal{A}_{*}^{k}\mathcal{A}_{*}^{n}+4A^{(5)}_{(i|klm|j)n}\hat{\mathcal{A}}^{ln}\mathcal{A}_{*}^{k}\mathcal{A}_{*}^{m},
\end{align}
\begin{align}
\label{Eq: Appendix LHS coefficients V}
\mathsf{D}_{ij}{}^{kl}&=F\delta_{(i}{}^{k}\delta_{j)}{}^{l}+2F_{X^{ij}}\hat{\mathcal{A}}^{kl}+A^{(1)}_{mn}\mathcal{A}_{*}^{m}\mathcal{A}_{*}^{n}\delta_{(i}{}^{k}\delta_{j)}{}^{l}-2A^{(1)}_{(i|m|}\hat{\mathcal{A}}^{(k|m|}\delta_{j)}{}^{l)}+\frac{1}{2}A^{(3)}_{ijmn}\hat{\mathcal{A}}^{kl}\mathcal{A}_{*}^{m}\mathcal{A}_{*}^{n}\nonumber\\&-A^{(3)}_{(i|m|j)n}\hat{\mathcal{A}}^{kl}\mathcal{A}_{*}^{m}\mathcal{A}_{*}^{n}-A^{(4)}_{(i|m|j)n}\hat{\mathcal{A}}^{(k|m|}\hat{\mathcal{A}}^{l)n}+A^{(4)}_{(i|mno|}\hat{\mathcal{A}}^{(k|m}\mathcal{A}_{*}^{n}\mathcal{A}_{*}^{o|}\delta_{j)}{}^{l)}\nonumber\\&-A^{(4)}_{(i|mno|}\hat{\mathcal{A}}^{(k|o}\mathcal{A}_{*}^{m}\mathcal{A}_{*}^{n|}\delta_{j)}{}^{l)}-A^{(4)}_{mn(i|o|}\hat{\mathcal{A}}^{(k|n}\mathcal{A}_{*}^{m}\mathcal{A}_{*}^{o|}\delta_{j)}{}^{l)}+A^{(4)}_{mn(i|o|}\hat{\mathcal{A}}^{(k|o}\mathcal{A}_{*}^{m}\mathcal{A}_{*}^{n|}\delta_{j)}{}^{l)}\nonumber\\&+A^{(5)}_{ijmnop}\hat{\mathcal{A}}^{km}\hat{\mathcal{A}}^{ln}\mathcal{A}_{*}^{o}\mathcal{A}_{*}^{p}-2A^{(5)}_{ijmnop}\hat{\mathcal{A}}^{(k|m|}\hat{\mathcal{A}}^{l)p}\mathcal{A}_{*}^{n}\mathcal{A}_{*}^{o}-2A^{(5)}_{(i|mno|j)p}\hat{\mathcal{A}}^{kn}\hat{\mathcal{A}}^{lo}\mathcal{A}_{*}^{m}\mathcal{A}_{*}^{p}\nonumber\\&+4A^{(5)}_{(i|mno|j)p}\hat{\mathcal{A}}^{(k|n|}\hat{\mathcal{A}}^{l)p}\mathcal{A}_{*}^{m}\mathcal{A}_{*}^{o}.
\end{align}

The coefficients $\mathcal{Z}_{i}$ and $\mathcal{Z}_{ijk}$ in the decomposition \eqref{Eq: RHS decomposition} are
\begin{align}
\label{Eq: Appendix RHS coefficients I}
\mathcal{Z}_{i}=2F_{X^{ij}}\mathcal{A}_{*}^{j}+A^{(2)}_{ij}\mathcal{A}_{*}^{j}-\frac{1}{2}A^{(3)}_{jkil}\mathcal{A}_{*}^{j}\mathcal{A}_{*}^{k}\mathcal{A}_{*}^{l},
\end{align}
\begin{align}
\mathcal{Z}_{ijk}&=\frac{1}{2}A^{(3)}_{jkli}\mathcal{A}_{*}^{l}-A^{(3)}_{(j|l|k)i}\mathcal{A}_{*}^{l}-A^{(4)}_{(j|l|k)i}\mathcal{A}_{*}^{l}-A^{(5)}_{jklmni}\mathcal{A}_{*}^{l}\mathcal{A}_{*}^{m}\mathcal{A}_{*}^{n}+2A^{(5)}_{(j|lmn|k)i}\mathcal{A}_{*}^{l}\mathcal{A}_{*}^{m}\mathcal{A}_{*}^{n}.\nonumber\\&
\label{Eq: Appendix RHS coefficients II}
\end{align}

With these blocks, the coefficient system \eqref{Eq: Auxiliary tensor solution I} determines $\alpha_{i}$ and $\beta_{ijk}$. The contractions entering the Schur complement are
\begin{align}
\label{Eq: Appendix Schur complement contraction I}
\mathcal{B}_{i}{}^{ab}h_{ab}&=6F_{X^{ij}}\mathcal{A}_{*}^{j}+3A^{(2)}_{ij}\mathcal{A}_{*}^{j}+\frac{1}{2}A^{(3)}_{jkli}\hat{\mathcal{A}}^{jk}\mathcal{A}_{*}^{l}-A^{(3)}_{ljki}\hat{\mathcal{A}}^{jk}\mathcal{A}_{*}^{l}-\frac{3}{2}A^{(3)}_{jkil}\mathcal{A}_{*}^{j}\mathcal{A}_{*}^{k}\mathcal{A}_{*}^{l}\nonumber\\&-A^{(4)}_{ljik}\hat{\mathcal{A}}^{jk}\mathcal{A}_{*}^{l}+2A^{(5)}_{jklmin}\hat{\mathcal{A}}^{mn}\mathcal{A}_{*}^{j}\mathcal{A}_{*}^{k}\mathcal{A}_{*}^{l}-A^{(5)}_{jkmnil}\hat{\mathcal{A}}^{mn}\mathcal{A}_{*}^{j}\mathcal{A}_{*}^{k}\mathcal{A}_{*}^{l},
\end{align}
\begin{align}
\label{Eq: Appendix Schur complement contraction II}
\mathcal{B}_{i}{}^{ab}\hat{\mathcal{A}}^{k}{}_{a}\hat{\mathcal{A}}^{l}{}_{b}&=2F_{X^{ij}}\hat{\mathcal{A}}^{kl}\mathcal{A}_{*}^{j}+A^{(2)}_{ij}\hat{\mathcal{A}}^{kl}\mathcal{A}_{*}^{j}-\frac{1}{2}A^{(3)}_{jmin}\hat{\mathcal{A}}^{kl}\mathcal{A}_{*}^{j}\mathcal{A}_{*}^{m}\mathcal{A}_{*}^{n}+\frac{1}{2}A^{(3)}_{jmni}\hat{\mathcal{A}}^{kj}\hat{\mathcal{A}}^{lm}\mathcal{A}_{*}^{n}\nonumber\\&-A^{(3)}_{jnmi}\hat{\mathcal{A}}^{(k|j|}\hat{\mathcal{A}}^{l)m}\mathcal{A}_{*}^{n}-A^{(4)}_{njim}\hat{\mathcal{A}}^{(k|j|}\hat{\mathcal{A}}^{l)m}\mathcal{A}_{*}^{n}-A^{(5)}_{nojmip}\hat{\mathcal{A}}^{kj}\hat{\mathcal{A}}^{lm}\mathcal{A}_{*}^{n}\mathcal{A}_{*}^{o}\mathcal{A}_{*}^{p}\nonumber\\&+2A^{(5)}_{nojpim}\hat{\mathcal{A}}^{(k|j|}\hat{\mathcal{A}}^{l)m}\mathcal{A}_{*}^{n}\mathcal{A}_{*}^{o}\mathcal{A}_{*}^{p}.
\end{align}

Therefore the Schur complement is
\begin{align}
\label{Eq: Appendix Schur complement expanded}
\mathcal{S}_{ij}&=A^{(1)}_{ij}+A^{(2)}_{ij}-A^{(3)}_{kl(ij)}\mathcal{A}_{*}^{k}\mathcal{A}_{*}^{l}-A^{(4)}_{klij}\mathcal{A}_{*}^{k}\mathcal{A}_{*}^{l}+A^{(5)}_{klmnij}\mathcal{A}_{*}^{k}\mathcal{A}_{*}^{l}\mathcal{A}_{*}^{m}\mathcal{A}_{*}^{n}\nonumber\\&-\Big(6\mathcal{A}_{*}^{k}F_{X^{ik}}+3A^{(2)}_{ik}\mathcal{A}_{*}^{k}-A^{(3)}_{klmi}\hat{\mathcal{A}}^{lm}\mathcal{A}_{*}^{k}+\frac{1}{2}A^{(3)}_{lmki}\hat{\mathcal{A}}^{lm}\mathcal{A}_{*}^{k}-\frac{3}{2}A^{(3)}_{klim}\mathcal{A}_{*}^{k}\mathcal{A}_{*}^{l}\mathcal{A}_{*}^{m}\nonumber\\&-A^{(4)}_{klim}\hat{\mathcal{A}}^{lm}\mathcal{A}_{*}^{k}+2A^{(5)}_{klmnio}\hat{\mathcal{A}}^{no}\mathcal{A}_{*}^{k}\mathcal{A}_{*}^{l}\mathcal{A}_{*}^{m}-A^{(5)}_{klnoim}\hat{\mathcal{A}}^{no}\mathcal{A}_{*}^{k}\mathcal{A}_{*}^{l}\mathcal{A}_{*}^{m}\Big)\alpha_{j}\nonumber\\&-\Big(2\hat{\mathcal{A}}^{kl}\mathcal{A}_{*}^{m}F_{X^{im}}+A^{(2)}_{im}\hat{\mathcal{A}}^{kl}\mathcal{A}_{*}^{m}-\frac{1}{2}A^{(3)}_{mnio}\hat{\mathcal{A}}^{kl}\mathcal{A}_{*}^{m}\mathcal{A}_{*}^{n}\mathcal{A}_{*}^{o}+\frac{1}{2}A^{(3)}_{mnoi}\hat{\mathcal{A}}^{km}\hat{\mathcal{A}}^{ln}\mathcal{A}_{*}^{o}\nonumber\\&-A^{(3)}_{mnoi}\hat{\mathcal{A}}^{km}\hat{\mathcal{A}}^{lo}\mathcal{A}_{*}^{n}-A^{(4)}_{mnio}\hat{\mathcal{A}}^{kn}\hat{\mathcal{A}}^{lo}\mathcal{A}_{*}^{m}-A^{(5)}_{mnopiq}\hat{\mathcal{A}}^{ko}\hat{\mathcal{A}}^{lp}\mathcal{A}_{*}^{m}\mathcal{A}_{*}^{n}\mathcal{A}_{*}^{q}\nonumber\\&+2A^{(5)}_{mnopiq}\hat{\mathcal{A}}^{ko}\hat{\mathcal{A}}^{lq}\mathcal{A}_{*}^{m}\mathcal{A}_{*}^{n}\mathcal{A}_{*}^{p}\Big)\beta_{jkl}.
\end{align}

The first line of Eq. \eqref{Eq: Appendix Schur complement expanded} is the scalar kinetic block $\mathcal{G}_{ij}$ of appendix \ref{AppSec: Explicit coefficients in the ADM action}, while the remaining terms give $-\mathcal{B}_{i}{}^{ab}\mathcal{W}_{jab}$. The displayed expression is not manifestly symmetric in $i,j$; the symmetry $\mathcal{S}_{ij}=\mathcal{S}_{ji}$ is realized once $\alpha_{j}$ and $\beta_{jkl}$ solve Eq. \eqref{Eq: Auxiliary tensor solution I}.
\section{Single-field limit}
\label{AppSec: Single-field limit}
This appendix collects the technical details of the single-field reduction. The first subsection is used in section \ref{Sec: Complete degeneracy in the metric-invertible branch}, where we check that the complete degeneracy condition reduces correctly when $\mathcal{N}=1$. The second subsection is used later in section \ref{Sec: Consistency checks and special subclasses}, after the Hamiltonian analysis of section \ref{Sec: Hamiltonian analysis}, to verify the primary-primary consistency condition in the same single-field limit.
\subsection{Reduction of the primary degeneracy condition}
\label{AppSubsec: Reduction of the primary degeneracy condition}
In this subsection we record the single-field reduction of the complete degeneracy condition derived in section \ref{Sec: Complete degeneracy in the metric-invertible branch}. When $\mathcal{N}=1$, all field-space indices can be dropped. The coefficient functions relevant for the velocity Hessian reduce as
\begin{align}
\label{Eq: Single-field reduction of coefficient functions}
&F\rightarrow F,\qquad
F_{X^{ij}}\rightarrow F_{X},\qquad
A^{(1)}_{ij}\rightarrow A^{(1)},\qquad
A^{(2)}_{ij}\rightarrow A^{(2)},\qquad
A^{(3)}_{ijkl}\rightarrow A^{(3)},\nonumber\\&
A^{(4)}_{ijkl}\rightarrow A^{(4)},\qquad
A^{(5)}_{ijklmn}\rightarrow A^{(5)}.
\end{align}

The lower-derivative coefficients and the terms linear in second derivatives do not enter the velocity Hessian, and therefore do not affect the primary degeneracy condition. The ADM variables reduce as
\begin{align}
\label{Eq: Single-field reduction of ADM variables}
\mathcal{A}_{*}^{i}\rightarrow\mathcal{A}_{*},\qquad
\hat{\mathcal{A}}^{i}{}_{a}\rightarrow\hat{\mathcal{A}}_{a},\qquad
\hat{\mathcal{A}}^{ij}\rightarrow\hat{\mathcal{A}}^{2},\qquad
X^{ij}\rightarrow X,
\end{align}
with
\begin{align}
\label{Eq: Single-field relation between X and spatial gradient}
\hat{\mathcal{A}}^{2}=X+\mathcal{A}_{*}^{2}.
\end{align}

The auxiliary tensor introduced in Eq. \eqref{Eq: Auxiliary tensor decomposition} takes the single-field form \eqref{Eq: Single-field auxiliary tensor in special section}. The source coefficients in Eq. \eqref{Eq: RHS decomposition} reduce to
\begin{align}
\label{Eq: Single-field source coefficients}
&\mathcal{Z}_{i}\rightarrow\mathcal{Z}_{h}=A^{(2)}\mathcal{A}_{*}-\frac{1}{2}A^{(3)}\mathcal{A}_{*}^{3}+2\mathcal{A}_{*}F_{X},\nonumber\\&
\mathcal{Z}_{ijk}\rightarrow\mathcal{Z}_{\hat{\mathcal{A}}}=-\frac{1}{2}\left(A^{(3)}+2A^{(4)}\right)\mathcal{A}_{*}+A^{(5)}\mathcal{A}_{*}^{3}.
\end{align}

The blocks of the coefficient matrix in Eq. \eqref{Eq: Auxiliary tensor solution II} reduce to
\begin{align}
\label{Eq: Single-field coefficient matrix blocks}
&\mathsf{A}=\frac{1}{2}\left(2A^{(1)}+6A^{(2)}-A^{(3)}\hat{\mathcal{A}}^{2}\right)\mathcal{A}_{*}^{2}-2F+2\hat{\mathcal{A}}^{2}F_{X},\nonumber\\&
\mathsf{B}=-\frac{1}{2}\hat{\mathcal{A}}^{2}\left(-2A^{(2)}\mathcal{A}_{*}^{2}+A^{(3)}\hat{\mathcal{A}}^{2}\mathcal{A}_{*}^{2}+2F-4\hat{\mathcal{A}}^{2}F_{X}\right),\nonumber\\&
\mathsf{C}=-2A^{(1)}-A^{(4)}\hat{\mathcal{A}}^{2}-\frac{3}{2}A^{(3)}\mathcal{A}_{*}^{2}+A^{(5)}\hat{\mathcal{A}}^{2}\mathcal{A}_{*}^{2}+6F_{X},\nonumber\\&
\mathsf{D}=A^{(1)}\left(-2\hat{\mathcal{A}}^{2}+\mathcal{A}_{*}^{2}\right)+F-\frac{1}{2}\hat{\mathcal{A}}^{2}\left[2A^{(4)}\hat{\mathcal{A}}^{2}+\left(A^{(3)}-2A^{(5)}\hat{\mathcal{A}}^{2}\right)\mathcal{A}_{*}^{2}-4F_{X}\right].
\end{align}

Thus the coefficient system \eqref{Eq: Auxiliary tensor solution II} becomes the two-dimensional linear system
\begin{align}
\label{Eq: Single-field auxiliary coefficient system}
\begin{pmatrix}
\mathsf{A}&\mathsf{B}\\
\mathsf{C}&\mathsf{D}
\end{pmatrix}
\begin{pmatrix}
\alpha\\
\beta
\end{pmatrix}
=
\begin{pmatrix}
\mathcal{Z}_{h}\\
\mathcal{Z}_{\hat{\mathcal{A}}}
\end{pmatrix}.
\end{align}

Solving Eq. \eqref{Eq: Single-field auxiliary coefficient system} gives
\begin{align}
\label{Eq: Single-field auxiliary tensor solution}
\alpha=\frac{\mathsf{D}\mathcal{Z}_{h}-\mathsf{B}\mathcal{Z}_{\hat{\mathcal{A}}}}{\mathsf{A}\mathsf{D}-\mathsf{B}\mathsf{C}},\qquad
\beta=\frac{-\mathsf{C}\mathcal{Z}_{h}+\mathsf{A}\mathcal{Z}_{\hat{\mathcal{A}}}}{\mathsf{A}\mathsf{D}-\mathsf{B}\mathsf{C}}.
\end{align}

In the regular metric-invertible branch, the determinant $\mathsf{A}\mathsf{D}-\mathsf{B}\mathsf{C}$ of the system \eqref{Eq: Single-field auxiliary coefficient system} is nonzero. The degeneracy condition is therefore obtained by setting the numerator of the resulting Schur complement to zero.

The single-field form of Eq. \eqref{Eq: Schur complement II} is
\begin{align}
\label{Eq: Single-field Schur complement before solving}
\mathcal{S}&=A^{(1)}+A^{(2)}-\left(A^{(3)}+A^{(4)}\right)\mathcal{A}_{*}^{2}+A^{(5)}\mathcal{A}_{*}^{4}\nonumber\\&+\frac{1}{2}\mathcal{A}_{*}\left[-6A^{(2)}+\left(A^{(3)}+2A^{(4)}\right)\hat{\mathcal{A}}^{2}+\left(3A^{(3)}-2A^{(5)}\hat{\mathcal{A}}^{2}\right)\mathcal{A}_{*}^{2}-12F_{X}\right]\alpha\nonumber\\&+\frac{1}{2}\hat{\mathcal{A}}^{2}\mathcal{A}_{*}\left[-2A^{(2)}+\left(A^{(3)}+2A^{(4)}\right)\hat{\mathcal{A}}^{2}+\left(A^{(3)}-2A^{(5)}\hat{\mathcal{A}}^{2}\right)\mathcal{A}_{*}^{2}-4F_{X}\right]\beta.
\end{align}

Solving Eq. \eqref{Eq: Single-field auxiliary coefficient system}, substituting the result into Eq. \eqref{Eq: Single-field Schur complement before solving}, and using Eq. \eqref{Eq: Single-field relation between X and spatial gradient}, the numerator of $\mathcal{S}$ can be written, up to an overall nonzero numerical factor, in the polynomial form
\begin{align}
\label{Eq: Single-field polynomial degeneracy condition}
0=\mathcal{P}_{0}+\mathcal{A}_{*}^{2}\mathcal{P}_{2}+\mathcal{A}_{*}^{4}\mathcal{P}_{4}.
\end{align}

The coefficient independent of $\mathcal{A}_{*}$ is
\begin{align}
\label{Eq: Single-field polynomial coefficient zero}
\mathcal{P}_{0}=-4\left(A^{(1)}+A^{(2)}\right)\left[-2F^{2}-8F_{X}^{2}X^{2}+FX\left(2A^{(1)}+4F_{X}+A^{(4)}X\right)\right].
\end{align}

The coefficient of $\mathcal{A}_{*}^{2}$ is
\begin{align}
\label{Eq: Single-field polynomial coefficient two}
\mathcal{P}_{2}&=-8F\left[\left(A^{(3)}+A^{(4)}\right)F-4A^{(2)}F_{X}-6F_{X}^{2}\right]+8A^{(1)3}X+8\Big(-2A^{(2)}A^{(4)}F\nonumber\\&+A^{(3)}FF_{X}+8A^{(2)}F_{X}^{2}\Big)X-\left(A^{(3)2}-4A^{(2)}A^{(5)}\right)FX^{2}-4A^{(1)}\Big[4F\left(A^{(2)}+F_{X}\right)\nonumber\\&+\left(-3A^{(3)}F+20A^{(2)}F_{X}\right)X-\left(3A^{(2)}A^{(4)}+A^{(5)}F-4A^{(3)}F_{X}\right)X^{2}\Big]-4A^{(1)2}\Big[F\nonumber\\&-X\left(6A^{(2)}-4F_{X}+A^{(4)}X\right)\Big].
\end{align}

The coefficient of $\mathcal{A}_{*}^{4}$ is
\begin{align}
\label{Eq: Single-field polynomial coefficient four}
\mathcal{P}_{4}&=4A^{(1)3}+A^{(1)}\Big[8A^{(3)}F+16F_{X}\left(-2A^{(2)}+F_{X}\right)+8A^{(3)}F_{X}X+3\Big(A^{(3)2}\nonumber\\&-4A^{(2)}A^{(5)}\Big)X^{2}\Big]+4\left[-4A^{(3)}FF_{X}+8A^{(2)}F_{X}^{2}-A^{(3)2}FX+2A^{(5)}F\left(F+2A^{(2)}X\right)\right]\nonumber\\&+4A^{(1)2}\left[2A^{(2)}-4F_{X}-X\left(A^{(3)}+A^{(5)}X\right)\right].
\end{align}

Because the coefficient functions depend on the fields through $X$, expressing the numerator in terms of $X$ and $\mathcal{A}_{*}$ by using Eq. \eqref{Eq: Single-field relation between X and spatial gradient} makes $F$, $F_{X}$ and $A^{(I)}$ independent of $\mathcal{A}_{*}$. At fixed $X$, the normal component $\mathcal{A}_{*}$ is unconstrained. Therefore the degeneracy condition holds for all $\mathcal{A}_{*}$ only if each coefficient in Eq. \eqref{Eq: Single-field polynomial degeneracy condition} vanishes separately:
\begin{align}
\label{Eq: Single-field degeneracy conditions}
\mathcal{P}_{0}=0,\qquad
\mathcal{P}_{2}=0,\qquad
\mathcal{P}_{4}=0.
\end{align}

These are the single-field conditions obtained from the complete degeneracy condition $\mathcal{S}=0$ in the metric-invertible branch. They take the form of a set of polynomial conditions on the coefficient functions, of the same type as those found in the degeneracy analysis of single-field quadratic higher-order scalar-tensor theories \cite{Langlois:2015skt}. This confirms that the multi-field construction of section \ref{Sec: Complete degeneracy in the metric-invertible branch} reduces consistently when $\mathcal{N}=1$.
\subsection{Primary-primary consistency in the single-field limit}
\label{AppSubsec: Primary-primary consistency in the single-field limit}
We now use the single-field reduction of appendix \ref{AppSubsec: Reduction of the primary degeneracy condition} to check, in the single-field limit, the primary-primary consistency condition of the Hamiltonian analysis of section \ref{Sec: Hamiltonian analysis}.

When $\mathcal{N}=1$, the non-derivative part of the primary-primary bracket has no nontrivial field-space antisymmetrization. Therefore the single-field reductions of Eqs. \eqref{Eq: Momentum dependent bracket coefficient} and \eqref{Eq: Momentum independent bracket coefficient} give
\begin{align}
\label{Eq: Single-field M and N vanish in appendix}
\mathfrak{M}=0,\qquad
\mathfrak{N}=0.
\end{align}

However, the derivative-smearing part of the bracket is not removed by this argument. The single-field version of Eq. \eqref{Eq: Derivative smearing coefficient} is
\begin{align}
\label{Eq: Single-field derivative smearing in appendix}
\mathfrak{D}^{a}=\sqrt{h}\left(\mathcal{J}^{a}+2\mathcal{W}_{bc}\mathcal{R}^{bca}\right).
\end{align}

The derivative decomposition of $\Delta$ is
\begin{align}
\label{Eq: Single-field Delta derivative decomposition}
\Delta=\Lambda+\mathcal{J}^{a}D_{a}\mathcal{A}_{*}+\mathcal{I}^{ab}S_{ab}.
\end{align}

By spatial covariance, the derivative coefficients can be decomposed as
\begin{align}
\label{Eq: Single-field J and I decomposition}
\mathcal{J}^{a}=\mathcal{J}_{\parallel}\hat{\mathcal{A}}^{a},\qquad
\mathcal{I}^{ab}=\mathcal{I}_{h}h^{ab}+\mathcal{I}_{\hat{\mathcal{A}}}\hat{\mathcal{A}}^{a}\hat{\mathcal{A}}^{b}.
\end{align}

Using Eq. \eqref{Eq: R tensor definition}, one obtains
\begin{align}
\label{Eq: Single-field WR contraction}
\mathcal{W}_{bc}\mathcal{R}^{bca}=\frac{1}{2}\hat{\mathcal{A}}^{a}\left[\alpha\left(\hat{\mathcal{A}}^{2}\mathcal{I}_{\hat{\mathcal{A}}}-\mathcal{I}_{h}\right)+\beta\left(\hat{\mathcal{A}}^{4}\mathcal{I}_{\hat{\mathcal{A}}}+\hat{\mathcal{A}}^{2}\mathcal{I}_{h}\right)\right].
\end{align}

Therefore
\begin{align}
\label{Eq: Single-field D explicit form}
\mathfrak{D}^{a}=\sqrt{h}\hat{\mathcal{A}}^{a}\left[\mathcal{J}_{\parallel}+\alpha\left(\hat{\mathcal{A}}^{2}\mathcal{I}_{\hat{\mathcal{A}}}-\mathcal{I}_{h}\right)+\beta\left(\hat{\mathcal{A}}^{4}\mathcal{I}_{\hat{\mathcal{A}}}+\hat{\mathcal{A}}^{2}\mathcal{I}_{h}\right)\right].
\end{align}

A direct substitution of the single-field reductions of $\mathcal{J}^{a}$ and $\mathcal{I}^{ab}$ from appendix \ref{AppSec: Explicit coefficients in the decomposition of Delta}, together with the solution \eqref{Eq: Single-field auxiliary tensor solution}, shows that the bracket in Eq. \eqref{Eq: Single-field D explicit form} vanishes once the single-field degeneracy conditions \eqref{Eq: Single-field degeneracy conditions} are imposed:
\begin{align}
\label{Eq: Single-field D bracket vanishes}
\left[\mathcal{J}_{\parallel}+\alpha\left(\hat{\mathcal{A}}^{2}\mathcal{I}_{\hat{\mathcal{A}}}-\mathcal{I}_{h}\right)+\beta\left(\hat{\mathcal{A}}^{4}\mathcal{I}_{\hat{\mathcal{A}}}+\hat{\mathcal{A}}^{2}\mathcal{I}_{h}\right)\right]_{\mathcal{P}_{0}=\mathcal{P}_{2}=\mathcal{P}_{4}=0}=0.
\end{align}

Hence
\begin{align}
\label{Eq: Single-field D final result}
\mathfrak{D}^{a}=0.
\end{align}

The single-field primary-primary bracket therefore vanishes in the regular branch with invertible metric block. The non-derivative part vanishes because there is no nontrivial field-space antisymmetrization. The derivative-smearing coefficient vanishes by Eq. \eqref{Eq: Single-field D final result}. Thus no additional primary-primary bracket condition arises in the single-field limit beyond the primary degeneracy condition.
\section{Explicit coefficients in the decomposition of \texorpdfstring{$\Delta_{i}$}{Delta{i}}}
\label{AppSec: Explicit coefficients in the decomposition of Delta}
In this appendix we give the explicit expressions for the coefficients appearing in the decomposition \eqref{Eq: Hamiltonian Delta decomposition}, which we recall here,
\begin{align}
\label{Eq: Appendix Delta decomposition}
\Delta_{i}=\Lambda_{i}+\mathcal{J}_{ij}{}^{a}D_{a}\mathcal{A}_{*}^{j}+\mathcal{I}_{ij}{}^{ab}S^{j}{}_{ab}.
\end{align}

For the last coefficient it is convenient to display the lowered-index form
\begin{align}
\label{Eq: Appendix I lowered definition}
\mathcal{I}_{ijab}=h_{ac}h_{bd}\mathcal{I}_{ij}{}^{cd}.
\end{align}

The derivative-free coefficient $\Lambda_{i}$ is
\begin{align}
\label{Eq: Appendix Delta L coefficient}
\Lambda_{i}&=Q_{i}-\mathcal{A}_{*}^{j}\left(2F_{\phi^{j}}+Q_{j}\right)\left(3\alpha_{i}+\hat{\mathcal{A}}^{kl}\beta_{ikl}\right)\nonumber\\&-\mathcal{A}_{*}^{j}\left(\mathcal{A}_{*}^{k}B_{jki}+\hat{\mathcal{A}}^{kl}\left[\left(-2B_{jkl}+B_{klj}\right)\alpha_{i}+\hat{\mathcal{A}}^{mn}\left(-2B_{jkm}+B_{kmj}\right)\beta_{iln}\right]\right).
\end{align}

The coefficient $\mathcal{J}_{ij}{}^{a}$ multiplying $D_{a}\mathcal{A}_{*}^{j}$ is
\begin{align}
\label{Eq: Appendix Delta J coefficient}
\mathcal{J}_{ij}{}^{a}&=-2A^{(3)}_{klji}\hat{\mathcal{A}}^{ka}\mathcal{A}_{*}^{l}-2A^{(4)}_{lkij}\hat{\mathcal{A}}^{ka}\mathcal{A}_{*}^{l}+4A^{(5)}_{lmknij}\hat{\mathcal{A}}^{ka}\mathcal{A}_{*}^{l}\mathcal{A}_{*}^{m}\mathcal{A}_{*}^{n}+\Big(-12F_{X^{jk}}\hat{\mathcal{A}}^{ka}\nonumber\\&+4A^{(1)}_{jk}\hat{\mathcal{A}}^{ka}+6A^{(3)}_{kljm}\hat{\mathcal{A}}^{ka}\mathcal{A}_{*}^{l}\mathcal{A}_{*}^{m}+2A^{(4)}_{kljm}\hat{\mathcal{A}}^{ka}\hat{\mathcal{A}}^{lm}+2A^{(4)}_{lkjm}\hat{\mathcal{A}}^{ka}\mathcal{A}_{*}^{l}\mathcal{A}_{*}^{m}\nonumber\\&-2A^{(4)}_{lmjk}\hat{\mathcal{A}}^{ka}\mathcal{A}_{*}^{l}\mathcal{A}_{*}^{m}-8A^{(5)}_{klmnjo}\hat{\mathcal{A}}^{ka}\hat{\mathcal{A}}^{no}\mathcal{A}_{*}^{l}\mathcal{A}_{*}^{m}+4A^{(5)}_{klnojm}\hat{\mathcal{A}}^{ka}\hat{\mathcal{A}}^{no}\mathcal{A}_{*}^{l}\mathcal{A}_{*}^{m}\Big)\alpha_{i}\nonumber\\&+\Big(-4F_{X^{jn}}\hat{\mathcal{A}}^{na}\hat{\mathcal{A}}^{lm}+2A^{(1)}_{jn}\hat{\mathcal{A}}^{ma}\hat{\mathcal{A}}^{ln}+2A^{(1)}_{jn}\hat{\mathcal{A}}^{la}\hat{\mathcal{A}}^{mn}+2A^{(3)}_{nojp}\hat{\mathcal{A}}^{na}\hat{\mathcal{A}}^{lm}\mathcal{A}_{*}^{o}\mathcal{A}_{*}^{p}\nonumber\\&+A^{(4)}_{nojp}\hat{\mathcal{A}}^{na}\hat{\mathcal{A}}^{lp}\hat{\mathcal{A}}^{mo}+A^{(4)}_{nojp}\hat{\mathcal{A}}^{na}\hat{\mathcal{A}}^{lo}\hat{\mathcal{A}}^{mp}-A^{(4)}_{nojp}\hat{\mathcal{A}}^{ma}\hat{\mathcal{A}}^{lp}\mathcal{A}_{*}^{n}\mathcal{A}_{*}^{o}\nonumber\\&+A^{(4)}_{npjo}\hat{\mathcal{A}}^{ma}\hat{\mathcal{A}}^{lp}\mathcal{A}_{*}^{n}\mathcal{A}_{*}^{o}-A^{(4)}_{nojp}\hat{\mathcal{A}}^{la}\hat{\mathcal{A}}^{mp}\mathcal{A}_{*}^{n}\mathcal{A}_{*}^{o}+A^{(4)}_{npjo}\hat{\mathcal{A}}^{la}\hat{\mathcal{A}}^{mp}\mathcal{A}_{*}^{n}\mathcal{A}_{*}^{o}\nonumber\\&-4A^{(5)}_{nopqjr}\hat{\mathcal{A}}^{na}\hat{\mathcal{A}}^{lq}\hat{\mathcal{A}}^{mr}\mathcal{A}_{*}^{o}\mathcal{A}_{*}^{p}-4A^{(5)}_{noprjq}\hat{\mathcal{A}}^{na}\hat{\mathcal{A}}^{lq}\hat{\mathcal{A}}^{mr}\mathcal{A}_{*}^{o}\mathcal{A}_{*}^{p}\nonumber\\&+4A^{(5)}_{noqrjp}\hat{\mathcal{A}}^{na}\hat{\mathcal{A}}^{lq}\hat{\mathcal{A}}^{mr}\mathcal{A}_{*}^{o}\mathcal{A}_{*}^{p}\Big)\beta_{ilm}.
\end{align}

The lowered-index form of the coefficient of $S^{j}{}_{ab}$ is
\begin{align}
\label{Eq: Appendix Delta I coefficient}
\mathcal{I}_{ijab}&=2A^{(2)}_{ij}h_{ab}-A^{(3)}_{klij}\mathcal{A}_{*}^{k}\mathcal{A}_{*}^{l}h_{ab}+A^{(3)}_{klji}\hat{\mathcal{A}}^{k}{}_{a}\hat{\mathcal{A}}^{l}{}_{b}-2A^{(5)}_{klmnij}\hat{\mathcal{A}}^{m}{}_{a}\hat{\mathcal{A}}^{n}{}_{b}\mathcal{A}_{*}^{k}\mathcal{A}_{*}^{l}\nonumber\\&+\alpha_{i}\Big(-2A^{(1)}_{jk}\mathcal{A}_{*}^{k}h_{ab}-6A^{(2)}_{jk}\mathcal{A}_{*}^{k}h_{ab}+2A^{(3)}_{klmj}\hat{\mathcal{A}}^{lm}\mathcal{A}_{*}^{k}h_{ab}-A^{(3)}_{lmkj}\hat{\mathcal{A}}^{lm}\mathcal{A}_{*}^{k}h_{ab}\nonumber\\&-3A^{(3)}_{kljm}\hat{\mathcal{A}}^{k}{}_{a}\hat{\mathcal{A}}^{l}{}_{b}\mathcal{A}_{*}^{m}-A^{(4)}_{kljm}\hat{\mathcal{A}}^{k}{}_{(a}\hat{\mathcal{A}}^{l}{}_{b)}\mathcal{A}_{*}^{m}+A^{(4)}_{kljm}\hat{\mathcal{A}}^{k}{}_{(a}\hat{\mathcal{A}}^{m}{}_{b)}\mathcal{A}_{*}^{l}\nonumber\\&-A^{(4)}_{kljm}\hat{\mathcal{A}}^{l}{}_{(a}\hat{\mathcal{A}}^{k}{}_{b)}\mathcal{A}_{*}^{m}+A^{(4)}_{kljm}\hat{\mathcal{A}}^{m}{}_{(a}\hat{\mathcal{A}}^{k}{}_{b)}\mathcal{A}_{*}^{l}-2A^{(5)}_{klmnjo}\hat{\mathcal{A}}^{k}{}_{a}\hat{\mathcal{A}}^{l}{}_{b}\hat{\mathcal{A}}^{mn}\mathcal{A}_{*}^{o}\nonumber\\&+4A^{(5)}_{klmnjo}\hat{\mathcal{A}}^{k}{}_{a}\hat{\mathcal{A}}^{l}{}_{b}\hat{\mathcal{A}}^{mo}\mathcal{A}_{*}^{n}\Big)+\Big(-2A^{(1)}_{jn}\hat{\mathcal{A}}^{l}{}_{a}\hat{\mathcal{A}}^{m}{}_{b}\mathcal{A}_{*}^{n}-2A^{(2)}_{jn}\hat{\mathcal{A}}^{lm}\mathcal{A}_{*}^{n}h_{ab}\nonumber\\&-A^{(3)}_{nojp}\hat{\mathcal{A}}^{n}{}_{a}\hat{\mathcal{A}}^{o}{}_{b}\hat{\mathcal{A}}^{lm}\mathcal{A}_{*}^{p}-A^{(3)}_{nopj}\hat{\mathcal{A}}^{ln}\hat{\mathcal{A}}^{mo}\mathcal{A}_{*}^{p}h_{ab}+2A^{(3)}_{nopj}\hat{\mathcal{A}}^{ln}\hat{\mathcal{A}}^{mp}\mathcal{A}_{*}^{o}h_{ab}\nonumber\\&-2A^{(4)}_{nojp}\hat{\mathcal{A}}^{l}{}_{(a}\hat{\mathcal{A}}^{n}{}_{b)}\hat{\mathcal{A}}^{mo}\mathcal{A}_{*}^{p}+2A^{(4)}_{nojp}\hat{\mathcal{A}}^{l}{}_{(a}\hat{\mathcal{A}}^{n}{}_{b)}\hat{\mathcal{A}}^{mp}\mathcal{A}_{*}^{o}-2A^{(5)}_{nopqjr}\hat{\mathcal{A}}^{n}{}_{a}\hat{\mathcal{A}}^{o}{}_{b}\hat{\mathcal{A}}^{lp}\hat{\mathcal{A}}^{mq}\mathcal{A}_{*}^{r}\nonumber\\&+4A^{(5)}_{nopqjr}\hat{\mathcal{A}}^{n}{}_{a}\hat{\mathcal{A}}^{o}{}_{b}\hat{\mathcal{A}}^{lp}\hat{\mathcal{A}}^{mr}\mathcal{A}_{*}^{q}\Big)\beta_{ilm}.
\end{align}
\section{Organization of the smeared primary-primary bracket}
\label{AppSec: Organization of the smeared primary-primary bracket}
This appendix records how the smeared bracket \eqref{Eq: Smeared primary primary structure} and the coefficients \eqref{Eq: Derivative smearing coefficient}--\eqref{Eq: Momentum independent bracket coefficient} are assembled from the constraint \eqref{Eq: Primary degeneracy constraints}, giving the origin of each structure rather than a full term-by-term computation. We work on the reduced phase space of subsection \ref{Subsec: Auxiliary constraints and reduced phase space}, on which the second-class pair $(\chi^{i}{}_{a},\hat{p}_{i}{}^{a})$ has been eliminated. Since $\Psi_{i}$ contains neither $\hat{p}_{i}{}^{a}$ nor $p_{i}$, we have $\{\Psi[f],\hat{\mathcal{A}}^{i}{}_{a}\}=0$ and $\{\Psi[f],\phi^{i}\}=0$, so that $\{\Psi[f],\chi^{i}{}_{a}\}=0$ for $\chi^{i}{}_{a}=\hat{\mathcal{A}}^{i}{}_{a}-D_{a}\phi^{i}$. The auxiliary Dirac bracket of two smeared constraints therefore reduces to the Poisson bracket,
\begin{align}
\label{Eq: Primary primary Dirac equals Poisson}
\{\Psi[f],\Psi[g]\}_{D,{\rm{aux}}}=\{\Psi[f],\Psi[g]\}.
\end{align}

By the same absence of $p_{i}$ and $\hat{p}_{i}{}^{a}$ in $\Psi_{i}$, the pairs $(\phi^{i},p_{i})$ and $(\hat{\mathcal{A}}^{i}{}_{a},\hat{p}_{i}{}^{a})$ do not contribute, and only $(\mathcal{A}_{*}^{i},p_{*i})$ and $(h_{ab},\pi^{ab})$ remain,
\begin{align}
\label{Eq: Smeared bracket two sectors}
\{\Psi[f],\Psi[g]\}&=\int d^{3}x\bigg(\frac{\delta\Psi[f]}{\delta\mathcal{A}_{*}^{i}}\frac{\delta\Psi[g]}{\delta p_{*i}}-\frac{\delta\Psi[f]}{\delta p_{*i}}\frac{\delta\Psi[g]}{\delta\mathcal{A}_{*}^{i}}\nonumber\\&+\frac{\delta\Psi[f]}{\delta h_{ab}}\frac{\delta\Psi[g]}{\delta\pi^{ab}}-\frac{\delta\Psi[f]}{\delta\pi^{ab}}\frac{\delta\Psi[g]}{\delta h_{ab}}\bigg).
\end{align}

The dependence of $\Psi_{i}$ on $\hat{\mathcal{A}}^{i}{}_{a}$ enters Eq. \eqref{Eq: Smeared bracket two sectors} through $\mathcal{W}_{iab}$ and $\Delta_{i}$, in particular through the contractions $\hat{\mathcal{A}}^{ij}=h^{ab}\hat{\mathcal{A}}^{i}{}_{a}\hat{\mathcal{A}}^{j}{}_{b}$ and $\hat{\mathcal{A}}^{ia}=h^{ab}\hat{\mathcal{A}}^{i}{}_{b}$, and through the spatial structure $S^{i}{}_{ab}=D_{(a}\hat{\mathcal{A}}^{i}{}_{b)}$. This is why the bracket involves algebraic derivatives with respect to $\mathcal{A}_{*}^{i}$ and $h_{ab}$, while the dependence on $\hat{\mathcal{A}}^{i}{}_{a}$ enters through these composite quantities rather than through a separate canonical variation.

The two momenta in $\Psi_{i}$ enter through
\begin{align}
\label{Eq: Constraint momentum derivatives}
\frac{\delta\Psi[g]}{\delta p_{*i}}=g^{i},\qquad
\frac{\delta\Psi[g]}{\delta\pi^{ab}}=-2g^{j}\mathcal{W}_{jab},
\end{align}
both at most linear in $\pi^{ab}$. Since $\Psi_{i}$ is itself linear in $\pi^{ab}$, the bracket is at most linear in $\pi^{ab}$, and the split $\mathfrak{C}_{ij}=\mathfrak{M}_{ij}+\mathfrak{N}_{ij}$ of Eq. \eqref{Eq: C decomposition} into a part linear in $\pi^{ab}$ and a part independent of it is exhaustive. The field-space antisymmetry $\mathfrak{C}_{ij}=-\mathfrak{C}_{ji}$ used in subsection \ref{Subsec: Preservation of the primary degeneracy constraints} is the antisymmetry of Eq. \eqref{Eq: Smeared bracket two sectors} under $f\leftrightarrow g$.

The variation $\delta\Psi[f]/\delta\mathcal{A}_{*}^{i}$ has an algebraic part, from the explicit $\mathcal{A}_{*}^{i}$-dependence of $\mathcal{W}_{jab}$ and $\Delta_{j}$, and a derivative part, from the term $\mathcal{J}_{jk}{}^{a}D_{a}\mathcal{A}_{*}^{k}$ in the decomposition \eqref{Eq: Hamiltonian Delta decomposition}; integrating the latter by parts produces one piece in which the derivative acts on the smearing function and one in which it acts on the coefficient. The variation $\delta\Psi[f]/\delta h_{ab}$ has three sources: the algebraic $h_{ab}$-dependence of $\mathcal{W}_{jab}$ and $\Delta_{j}$; the prefactor $\sqrt{h}$, with $\delta\sqrt{h}/\delta h_{ab}=\frac{1}{2}\sqrt{h}h^{ab}$; and the spatial connection inside $S^{j}{}_{ab}$, whose variation is collected in the tensor $\mathcal{R}_{i}{}^{abc}$ of Eq. \eqref{Eq: R tensor definition}.

Combining these variations through Eqs. \eqref{Eq: Smeared bracket two sectors} and \eqref{Eq: Constraint momentum derivatives} gives the three coefficients. The momentum-dependent coefficient $\mathfrak{M}_{ij}$ collects the terms that keep a factor of $\pi^{ab}$: the $\mathcal{A}_{*}^{i}$-variation of $\mathcal{W}_{jab}\pi^{ab}$ in the $(\mathcal{A}_{*}^{i},p_{*i})$ sector gives the $(\partial\mathcal{W}/\partial\mathcal{A}_{*})_{\rm{alg}}$ terms, and its $h_{ab}$-variation in the $(h_{ab},\pi^{ab})$ sector gives the $(\partial\mathcal{W}/\partial h)_{\rm{alg}}\mathcal{W}$ terms. The momentum-independent coefficient $\mathfrak{N}_{ij}$ collects the rest: the algebraic terms from the $\mathcal{A}_{*}^{i}$- and $h_{ab}$-variations of $\Delta_{j}$; the trace term $2\mathcal{W}_{[i|a|}{}^{a}\Delta_{j]}$ from the $h^{ab}$ in $\delta\sqrt{h}/\delta h_{ab}$; the term $-2D_{a}\mathcal{J}_{[ij]}{}^{a}$, which is the piece of the $\mathcal{J}_{jk}{}^{a}D_{a}\mathcal{A}_{*}^{k}$ variation with the derivative on the coefficient; and the term $4\mathcal{W}_{[i|ab|}D_{c}\mathcal{R}_{j]}{}^{abc}$, the corresponding piece of the connection variation. The complementary pieces, with the derivative on the smearing function, are kept in $\mathfrak{D}_{ij}{}^{a}$: the $\mathcal{J}_{ji}{}^{a}$ term and the $2\mathcal{W}_{ibc}\mathcal{R}_{j}{}^{bca}$ term. This is the integration-by-parts convention noted below Eq. \eqref{Eq: Weighted antisymmetrization}: which of $\mathfrak{D}_{ij}{}^{a}$ and $\mathfrak{N}_{ij}$ carries a given $\mathcal{J}$ or $\mathcal{R}$ contribution depends on where the derivative is placed, while their sum as a differential operator on $f^{i}$ and $g^{j}$ does not.
\printbibliography
\end{document}